\documentclass[conference,10pt,twocolumn,letter]{IEEEtran}
\IEEEoverridecommandlockouts

\usepackage{cite}
\usepackage[T1]{fontenc}
\usepackage{graphicx}
\usepackage{amssymb}
\usepackage{amsmath}
\usepackage{amsthm}
\usepackage{enumerate}
\usepackage[shortlabels]{enumitem}
\usepackage{microtype}
\usepackage{balance}
\usepackage{subfigure}
\usepackage{xcolor} 
\usepackage{algorithm}
\usepackage{algorithmic}
\usepackage{amssymb}
\usepackage{amsfonts}
\usepackage{mathrsfs}
\usepackage{xspace}
\usepackage{bm}
\usepackage{upgreek}

\newcommand{\safemath}[2]{\newcommand{#1}{\ensuremath{#2}\xspace}}

\safemath{\bma}{\mathbf{a}}
\safemath{\bmb}{\mathbf{b}}
\safemath{\bmc}{\mathbf{c}}
\safemath{\bmd}{\mathbf{d}}
\safemath{\bme}{\mathbf{e}}
\safemath{\bmf}{\mathbf{f}}
\safemath{\bmg}{\mathbf{g}}
\safemath{\bmh}{\mathbf{h}}
\safemath{\bmi}{\mathbf{i}}
\safemath{\bmj}{\mathbf{j}}
\safemath{\bmk}{\mathbf{k}}
\safemath{\bml}{\mathbf{l}}
\safemath{\bmm}{\mathbf{m}}
\safemath{\bmn}{\mathbf{n}}
\safemath{\bmo}{\mathbf{o}}
\safemath{\bmp}{\mathbf{p}}
\safemath{\bmq}{\mathbf{q}}
\safemath{\bmr}{\mathbf{r}}
\safemath{\bms}{\mathbf{s}}
\safemath{\bmt}{\mathbf{t}}
\safemath{\bmu}{\mathbf{u}}
\safemath{\bmv}{\mathbf{v}}
\safemath{\bmw}{\mathbf{w}}
\safemath{\bmx}{\mathbf{x}}
\safemath{\bmy}{\mathbf{y}}
\safemath{\bmz}{\mathbf{z}}
\safemath{\bmzero}{\mathbf{0}}
\safemath{\bmone}{\mathbf{1}}

\bmdefine{\biad}{a}
\bmdefine{\bibd}{b}
\bmdefine{\bicd}{c}
\bmdefine{\bidd}{d}
\bmdefine{\bied}{e}
\bmdefine{\bifd}{f}
\bmdefine{\bigd}{g}
\bmdefine{\bihd}{h}
\bmdefine{\biid}{i}
\bmdefine{\bijd}{j}
\bmdefine{\bikd}{k}
\bmdefine{\bild}{l}
\bmdefine{\bimd}{m}
\bmdefine{\bind}{n}
\bmdefine{\biod}{o}
\bmdefine{\bipd}{p}
\bmdefine{\biqd}{q}
\bmdefine{\bird}{r}
\bmdefine{\bisd}{s}
\bmdefine{\bitd}{t}
\bmdefine{\biud}{u}
\bmdefine{\bivd}{v}
\bmdefine{\biwd}{w}
\bmdefine{\bixd}{x}
\bmdefine{\biyd}{y}
\bmdefine{\bizd}{z}

\bmdefine{\bixid}{\xi}
\bmdefine{\bilambdad}{\lambda}
\bmdefine{\bimud}{\mu}
\bmdefine{\bithetad}{\theta}
\bmdefine{\biphid}{\phi}
\bmdefine{\bideltad}{\delta}

\safemath{\bmia}{\biad}
\safemath{\bmib}{\bibd}
\safemath{\bmic}{\bicd}
\safemath{\bmid}{\bidd}
\safemath{\bmie}{\bied}
\safemath{\bmif}{\bifd}
\safemath{\bmig}{\bigd}
\safemath{\bmih}{\bihd}
\safemath{\bmii}{\biid}
\safemath{\bmij}{\bijd}
\safemath{\bmik}{\bikd}
\safemath{\bmil}{\bild}
\safemath{\bmim}{\bimd}
\safemath{\bmin}{\bind}
\safemath{\bmio}{\biod}
\safemath{\bmip}{\bipd}
\safemath{\bmiq}{\biqd}
\safemath{\bmir}{\bird}
\safemath{\bmis}{\bisd}
\safemath{\bmit}{\bitd}
\safemath{\bmiu}{\biud}
\safemath{\bmiv}{\bivd}
\safemath{\bmiw}{\biwd}
\safemath{\bmix}{\bixd}
\safemath{\bmiy}{\biyd}
\safemath{\bmiz}{\bizd}

\safemath{\bmxi}{\bixid}
\safemath{\bmlambda}{\bilambdad}
\safemath{\bmmu}{\bimud}
\safemath{\bmtheta}{\bithetad}
\safemath{\bmphi}{\biphid}
\safemath{\bmdelta}{\bideltad}

\safemath{\bA}{\mathbf{A}}
\safemath{\bB}{\mathbf{B}}
\safemath{\bC}{\mathbf{C}}
\safemath{\bD}{\mathbf{D}}
\safemath{\bE}{\mathbf{E}}
\safemath{\bF}{\mathbf{F}}
\safemath{\bG}{\mathbf{G}}
\safemath{\bH}{\mathbf{H}}
\safemath{\bI}{\mathbf{I}}
\safemath{\bJ}{\mathbf{J}}
\safemath{\bK}{\mathbf{K}}
\safemath{\bL}{\mathbf{L}}
\safemath{\bM}{\mathbf{M}}
\safemath{\bN}{\mathbf{N}}
\safemath{\bO}{\mathbf{O}}
\safemath{\bP}{\mathbf{P}}
\safemath{\bQ}{\mathbf{Q}}
\safemath{\bR}{\mathbf{R}}
\safemath{\bS}{\mathbf{S}}
\safemath{\bT}{\mathbf{T}}
\safemath{\bU}{\mathbf{U}}
\safemath{\bV}{\mathbf{V}}
\safemath{\bW}{\mathbf{W}}
\safemath{\bX}{\mathbf{X}}
\safemath{\bY}{\mathbf{Y}}
\safemath{\bZ}{\mathbf{Z}}

\safemath{\bZero}{\mathbf{0}}
\safemath{\bOne}{\mathbf{1}}
\safemath{\bDelta}{\mathbf{\Delta}}
\safemath{\bLambda}{\mathbf{\UpLambda}}
\safemath{\bPhi}{\mathbf{\Upphi}}
\safemath{\bSigma}{\mathbf{\Upsigma}}
\safemath{\bOmega}{\mathbf{\Upomega}}
\safemath{\bTheta}{\mathbf{\Uptheta}}

\bmdefine{\biAd}{A}
\bmdefine{\biBd}{B}
\bmdefine{\biCd}{C}
\bmdefine{\biDd}{D}
\bmdefine{\biEd}{E}
\bmdefine{\biFd}{F}
\bmdefine{\biGd}{G}
\bmdefine{\biHd}{H}
\bmdefine{\biId}{I}
\bmdefine{\biJd}{J}
\bmdefine{\biKd}{K}
\bmdefine{\biLd}{L}
\bmdefine{\biMd}{M}
\bmdefine{\biOd}{N}
\bmdefine{\biPd}{O}
\bmdefine{\biQd}{P}
\bmdefine{\biRd}{R}
\bmdefine{\biSd}{S}
\bmdefine{\biTd}{T}
\bmdefine{\biUd}{U}
\bmdefine{\biVd}{V}
\bmdefine{\biWd}{W}
\bmdefine{\biXd}{X}
\bmdefine{\biYd}{Y}
\bmdefine{\biZd}{Z}

\bmdefine{\biDelta}{\Delta}
\bmdefine{\biLambda}{\Lambda}
\bmdefine{\biPhi}{\Phi}
\bmdefine{\biSigma}{\Sigma}
\bmdefine{\biOmega}{\Omega}
\bmdefine{\biTheta}{\Theta}

\safemath{\bimA}{\biAd}
\safemath{\bimB}{\biBd}
\safemath{\bimC}{\biCd}
\safemath{\bimD}{\biDd}
\safemath{\bimE}{\biEd}
\safemath{\bimF}{\biFd}
\safemath{\bimG}{\biGd}
\safemath{\bimH}{\biHd}
\safemath{\bimI}{\biId}
\safemath{\bimJ}{\biJd}
\safemath{\bimK}{\biKd}
\safemath{\bimL}{\biLd}
\safemath{\bimM}{\biMd}
\safemath{\bimN}{\biNd}
\safemath{\bimO}{\biOd}
\safemath{\bimP}{\biPd}
\safemath{\bimQ}{\biQd}
\safemath{\bimR}{\biRd}
\safemath{\bimS}{\biSd}
\safemath{\bimT}{\biTd}
\safemath{\bimU}{\biUd}
\safemath{\bimV}{\biVd}
\safemath{\bimW}{\biWd}
\safemath{\bimX}{\biXd}
\safemath{\bimY}{\biYd}
\safemath{\bimZ}{\biZd}

\safemath{\bimDelta}{\biDelta}
\safemath{\bimLambda}{\biLambda}
\safemath{\bimPhi}{\biPhi}
\safemath{\bimSigma}{\biSigma}
\safemath{\bimOmega}{\biOmega}
\safemath{\bimTheta}{\biTheta}

\safemath{\setA}{\mathcal{A}}
\safemath{\setB}{\mathcal{B}}
\safemath{\setC}{\mathcal{C}}
\safemath{\setD}{\mathcal{D}}
\safemath{\setE}{\mathcal{E}}
\safemath{\setF}{\mathcal{F}}
\safemath{\setG}{\mathcal{G}}
\safemath{\setH}{\mathcal{H}}
\safemath{\setI}{\mathcal{I}}
\safemath{\setJ}{\mathcal{J}}
\safemath{\setK}{\mathcal{K}}
\safemath{\setL}{\mathcal{L}}
\safemath{\setM}{\mathcal{M}}
\safemath{\setN}{\mathcal{N}}
\safemath{\setO}{\mathcal{O}}
\safemath{\setP}{\mathcal{P}}
\safemath{\setQ}{\mathcal{Q}}
\safemath{\setR}{\mathcal{R}}
\safemath{\setS}{\mathcal{S}}
\safemath{\setT}{\mathcal{T}}
\safemath{\setU}{\mathcal{U}}
\safemath{\setV}{\mathcal{V}}
\safemath{\setW}{\mathcal{W}}
\safemath{\setX}{\mathcal{X}}
\safemath{\setY}{\mathcal{Y}}
\safemath{\setZ}{\mathcal{Z}}
\safemath{\emptySet}{\varnothing}

\safemath{\colA}{\mathscr{A}}
\safemath{\colB}{\mathscr{B}}
\safemath{\colC}{\mathscr{C}}
\safemath{\colD}{\mathscr{D}}
\safemath{\colE}{\mathscr{E}}
\safemath{\colF}{\mathscr{F}}
\safemath{\colG}{\mathscr{G}}
\safemath{\colH}{\mathscr{H}}
\safemath{\colI}{\mathscr{I}}
\safemath{\colJ}{\mathscr{J}}
\safemath{\colK}{\mathscr{K}}
\safemath{\colL}{\mathscr{L}}
\safemath{\colM}{\mathscr{M}}
\safemath{\colN}{\mathscr{N}}
\safemath{\colO}{\mathscr{O}}
\safemath{\colP}{\mathscr{P}}
\safemath{\colQ}{\mathscr{Q}}
\safemath{\colR}{\mathscr{R}}
\safemath{\colS}{\mathscr{S}}
\safemath{\colT}{\mathscr{T}}
\safemath{\colU}{\mathscr{U}}
\safemath{\colV}{\mathscr{V}}
\safemath{\colW}{\mathscr{W}}
\safemath{\colX}{\mathscr{X}}
\safemath{\colY}{\mathscr{Y}}
\safemath{\colZ}{\mathscr{Z}}

\safemath{\opA}{\mathbb{A}}
\safemath{\opB}{\mathbb{B}}
\safemath{\opC}{\mathbb{C}}
\safemath{\opD}{\mathbb{D}}
\safemath{\opE}{\mathbb{E}}
\safemath{\opF}{\mathbb{F}}
\safemath{\opG}{\mathbb{G}}
\safemath{\opH}{\mathbb{H}}
\safemath{\opI}{\mathbb{I}}
\safemath{\opJ}{\mathbb{J}}
\safemath{\opK}{\mathbb{K}}
\safemath{\opL}{\mathbb{L}}
\safemath{\opM}{\mathbb{M}}
\safemath{\opN}{\mathbb{N}}
\safemath{\opO}{\mathbb{O}}
\safemath{\opP}{\mathbb{P}}
\safemath{\opQ}{\mathbb{Q}}
\safemath{\opR}{\mathbb{R}}
\safemath{\opS}{\mathbb{S}}
\safemath{\opT}{\mathbb{T}}
\safemath{\opU}{\mathbb{U}}
\safemath{\opV}{\mathbb{V}}
\safemath{\opW}{\mathbb{W}}
\safemath{\opX}{\mathbb{X}}
\safemath{\opY}{\mathbb{Y}}
\safemath{\opZ}{\mathbb{Z}}
\safemath{\opZero}{\mathbb{O}}
\safemath{\identityop}{\opI}

\safemath{\veca}{\bma}
\safemath{\vecb}{\bmb}
\safemath{\vecc}{\bmc}
\safemath{\vecd}{\bmd}
\safemath{\vece}{\bme}
\safemath{\vecf}{\bmf}
\safemath{\vecg}{\bmg}
\safemath{\vech}{\bmh}
\safemath{\veci}{\bmi}
\safemath{\vecj}{\bmj}
\safemath{\veck}{\bmk}
\safemath{\vecl}{\bml}
\safemath{\vecm}{\bmm}
\safemath{\vecn}{\bmn}
\safemath{\veco}{\bmo}
\safemath{\vecp}{\bmp}
\safemath{\vecq}{\bmq}
\safemath{\vecr}{\bmr}
\safemath{\vecs}{\bms}
\safemath{\vect}{\bmt}
\safemath{\vecu}{\bmu}
\safemath{\vecv}{\bmv}
\safemath{\vecw}{\bmw}
\safemath{\vecx}{\bmx}
\safemath{\vecy}{\bmy}
\safemath{\vecz}{\bmz}

\safemath{\veczero}{\bmzero}
\safemath{\vecone}{\bmone}
\safemath{\vecxi}{\bmxi}
\safemath{\veclambda}{\bmlambda}
\safemath{\vecmu}{\bmmu}
\safemath{\vectheta}{\bmtheta}
\safemath{\vecphi}{\bmphi}
\safemath{\vecdelta}{\bmdelta}

\safemath{\matA}{\bA}
\safemath{\matB}{\bB}
\safemath{\matC}{\bC}
\safemath{\matD}{\bD}
\safemath{\matE}{\bE}
\safemath{\matF}{\bF}
\safemath{\matG}{\bG}
\safemath{\matH}{\bH}
\safemath{\matI}{\bI}
\safemath{\matJ}{\bJ}
\safemath{\matK}{\bK}
\safemath{\matL}{\bL}
\safemath{\matM}{\bM}
\safemath{\matN}{\bN}
\safemath{\matO}{\bO}
\safemath{\matP}{\bP}
\safemath{\matQ}{\bQ}
\safemath{\matR}{\bR}
\safemath{\matS}{\bS}
\safemath{\matT}{\bT}
\safemath{\matU}{\bU}
\safemath{\matV}{\bV}
\safemath{\matW}{\bW}
\safemath{\matX}{\bX}
\safemath{\matY}{\bY}
\safemath{\matZ}{\bZ}
\safemath{\matzero}{\bmzero}

\safemath{\matDelta}{\bDelta}
\safemath{\matLambda}{\bLambda}
\safemath{\matPhi}{\bPhi}
\safemath{\matSigma}{\bSigma}
\safemath{\matOmega}{\bOmega}
\safemath{\matTheta}{\bTheta}

\safemath{\matidentity}{\matI}
\safemath{\matone}{\matO}

\safemath{\rnda}{A}
\safemath{\rndb}{B}
\safemath{\rndc}{C}
\safemath{\rndd}{D}
\safemath{\rnde}{E}
\safemath{\rndf}{F}
\safemath{\rndg}{G}
\safemath{\rndh}{H}
\safemath{\rndi}{I}
\safemath{\rndj}{J}
\safemath{\rndk}{K}
\safemath{\rndl}{L}
\safemath{\rndm}{M}
\safemath{\rndn}{N}
\safemath{\rndo}{O}
\safemath{\rndp}{P}
\safemath{\rndq}{Q}
\safemath{\rndr}{R}
\safemath{\rnds}{S}
\safemath{\rndt}{T}
\safemath{\rndu}{U}
\safemath{\rndv}{V}
\safemath{\rndw}{W}
\safemath{\rndx}{X}
\safemath{\rndy}{Y}
\safemath{\rndz}{Z}

\safemath{\rveca}{\bimA}
\safemath{\rvecb}{\bimB}
\safemath{\rvecc}{\bimC}
\safemath{\rvecd}{\bimD}
\safemath{\rvece}{\bimE}
\safemath{\rvecf}{\bimF}
\safemath{\rvecg}{\bimG}
\safemath{\rvech}{\bimH}
\safemath{\rveci}{\bimI}
\safemath{\rvecj}{\bimJ}
\safemath{\rveck}{\bimK}
\safemath{\rvecl}{\bimL}
\safemath{\rvecm}{\bimM}
\safemath{\rvecn}{\bimN}
\safemath{\rveco}{\bomO}
\safemath{\rvecp}{\bimP}
\safemath{\rvecq}{\bimQ}
\safemath{\rvecr}{\bimR}
\safemath{\rvecs}{\bimS}
\safemath{\rvect}{\bimT}
\safemath{\rvecu}{\bimU}
\safemath{\rvecv}{\bimV}
\safemath{\rvecw}{\bimW}
\safemath{\rvecx}{\bimX}
\safemath{\rvecy}{\bimY}
\safemath{\rvecz}{\bimZ}

\safemath{\rvecxi}{\bmxi}
\safemath{\rveclambda}{\bmlambda}
\safemath{\rvecmu}{\bmmu}
\safemath{\rvectheta}{\bmtheta}
\safemath{\rvecphi}{\bmphi}

\safemath{\rmatA}{\bimA}
\safemath{\rmatB}{\bimB}
\safemath{\rmatC}{\bimC}
\safemath{\rmatD}{\bimD}
\safemath{\rmatE}{\bimE}
\safemath{\rmatF}{\bimF}
\safemath{\rmatG}{\bimG}
\safemath{\rmatH}{\bimH}
\safemath{\rmatI}{\bimI}
\safemath{\rmatJ}{\bimJ}
\safemath{\rmatK}{\bimK}
\safemath{\rmatL}{\bimL}
\safemath{\rmatM}{\bimM}
\safemath{\rmatN}{\bimN}
\safemath{\rmatO}{\bimO}
\safemath{\rmatP}{\bimP}
\safemath{\rmatQ}{\bimQ}
\safemath{\rmatR}{\bimR}
\safemath{\rmatS}{\bimS}
\safemath{\rmatT}{\bimT}
\safemath{\rmatU}{\bimU}
\safemath{\rmatV}{\bimV}
\safemath{\rmatW}{\bimW}
\safemath{\rmatX}{\bimX}
\safemath{\rmatY}{\bimY}
\safemath{\rmatZ}{\bimZ}

\safemath{\rmatDelta}{\bimDelta}
\safemath{\rmatLambda}{\bimLambda}
\safemath{\rmatPhi}{\bimPhi}
\safemath{\rmatSigma}{\bimSigma}
\safemath{\rmatOmega}{\bimOmega}
\safemath{\rmatTheta}{\bimTheta}

\usepackage{amssymb}
\usepackage{amsfonts}
\usepackage{mathrsfs}
\usepackage{xspace}
\usepackage{bm}
\usepackage{fancyref}
\usepackage{textcomp}

\usepackage{multirow}
\usepackage{stmaryrd}

\newenvironment{textbmatrix}{	\setlength{\arraycolsep}{2.5pt}%
								\big[\begin{matrix}}{\end{matrix}\big]%
								\raisebox{0.08ex}{\vphantom{M}}}

\def\be{\begin{equation}}
\def\ee{\end{equation}}
\def\een{\nonumber \end{equation}}
\def\mat{\begin{bmatrix}}
\def\emat{\end{bmatrix}}
\def\btm{\begin{textbmatrix}}
\def\etm{\end{textbmatrix}}

\def\ba#1\ea{\begin{align}#1\end{align}}
\def\bas#1\eas{\begin{align*}#1\end{align*}}
\def\bs#1\es{\begin{split}#1\end{split}}
\def\bg#1\eg{\begin{gather}#1\end{gather}}
\def\bml#1\eml{\begin{multline}#1\end{multline}}
\def\bi#1\ei{\begin{itemize}#1\end{itemize}}

\newcommand{\lefto}{\mathopen{}\left}

\DeclareMathOperator*{\argmin}{arg\;min}		

\newcommand{\vecnorm}[1]{\lefto\lVert#1\right\rVert}		

\safemath{\dirac}{\delta}					
\safemath{\krond}{\dirac}					

\safemath{\upto}{\uparrow}
\safemath{\downto}{\downarrow}
\safemath{\iu}{j}							
\safemath{\ev}{\lambda}						
\safemath{\hilseqspace}{l^{2}}				
\newcommand{\banachfunspace}[1]{\setL^{#1}}	
\safemath{\hilfunspace}{\banachfunspace{2}}	
\newcommand{\floor}[1]{\lfloor #1 \rfloor}

\safemath{\SNR}{\textit{SNR}} 				
\safemath{\PAR}{\textit{PAR}} 				
\safemath{\No}{N_0}							
\safemath{\Es}{E_s}							
\safemath{\Eb}{E_b}							
\safemath{\EbNo}{\frac{\Eb}{\No}}
\safemath{\EsNo}{\frac{\Es}{\No}}

\DeclareMathOperator{\CHop}{\ensuremath{\opH}} 
\safemath{\tvir}{\rndh_{\CHop}}				
\safemath{\tvtf}{\rndl_{\CHop}}				
\safemath{\spf}{\rnds_{\CHop}}				
\safemath{\bff}{H_{\CHop}}					

\safemath{\ircf}{r_{h}}						
\safemath{\tftvcf}{r_{s}}					
\safemath{\tfcf}{r_{l}}						
\safemath{\bfcf}{r_{H}}						

\safemath{\tcorr}{c_h}						
\safemath{\scf}{c_{s}}						
\safemath{\tfcorr}{c_{l}}					
\safemath{\fcorr}{c_{H}}						

\safemath{\mi}{I}							
\safemath{\capacity}{C}						

\safemath{\normal}{\mathcal{N}}			
\safemath{\jpg}{\mathcal{CN}}			
\safemath{\mchain}{\leftrightarrow}		

\safemath{\dB}{\,\mathrm{dB}}
\safemath{\dBm}{\,\mathrm{dBm}}
\safemath{\Hz}{\,\mathrm{Hz}}
\safemath{\kHz}{\,\mathrm{kHz}}
\safemath{\MHz}{\,\mathrm{MHz}}
\safemath{\GHz}{\,\mathrm{GHz}}
\safemath{\s}{\,\mathrm{s}}
\safemath{\ms}{\,\mathrm{ms}}
\safemath{\mus}{\,\mathrm{\text{\textmu}s}}
\safemath{\ns}{\,\mathrm{ns}}
\safemath{\ps}{\,\mathrm{ps}}
\safemath{\meter}{\,\mathrm{m}}
\safemath{\mm}{\,\mathrm{mm}}
\safemath{\cm}{\,\mathrm{cm}}
\safemath{\m}{\,\mathrm{m}}
\safemath{\W}{\,\mathrm{W}}
\safemath{\mW}{\, \mathrm{mW}}
\safemath{\J}{\,\mathrm{J}}
\safemath{\K}{\,\mathrm{K}}
\safemath{\bit}{\,\mathrm{bit}}
\safemath{\nat}{\,\mathrm{nat}}

\safemath{\define}{\triangleq}			

\safemath{\equivalent}{\sim}
\safemath{\distas}{\sim}					
\safemath{\sdiff}{\Delta}				

\safemath{\reals}{\mathbb{R}}
\safemath{\positivereals}{\reals_{+}}
\safemath{\integers}{\mathbb{Z}}
\safemath{\posint}{\integers_{+}}
\safemath{\naturals}{\mathbb{N}}
\safemath{\posnaturals}{\naturals_{+}}
\safemath{\complexset}{\mathbb{C}}
\safemath{\rationals}{\mathbb{Q}}

\newcommand*{\fancyrefapplabelprefix}{app}		
\newcommand*{\fancyrefthmlabelprefix}{thm}		
\newcommand*{\fancyreflemlabelprefix}{lem}		
\newcommand*{\fancyrefcorlabelprefix}{cor}		
\newcommand*{\fancyrefdeflabelprefix}{def}		
\newcommand*{\fancyrefproplabelprefix}{prop}		
\newcommand*{\fancyrefexmpllabelprefix}{exmpl}
\newcommand*{\fancyrefalglabelprefix}{alg}		
\newcommand*{\fancyreftbllabelprefix}{tbl}		

\frefformat{vario}{\fancyrefseclabelprefix}{Sec.~#1}
\frefformat{vario}{\fancyrefthmlabelprefix}{Thm.~#1}
\frefformat{vario}{\fancyreftbllabelprefix}{Tbl.~#1}
\frefformat{vario}{\fancyreflemlabelprefix}{Lem.~#1}
\frefformat{vario}{\fancyrefcorlabelprefix}{Cor.~#1}
\frefformat{vario}{\fancyrefdeflabelprefix}{Def.~#1}
\frefformat{vario}{\fancyreffiglabelprefix}{Fig.~#1}
\frefformat{vario}{\fancyrefapplabelprefix}{App.~#1}
\frefformat{vario}{\fancyrefeqlabelprefix}{(#1)}
\frefformat{vario}{\fancyrefproplabelprefix}{Prop.~#1}
\frefformat{vario}{\fancyrefexmpllabelprefix}{Ex.~#1}
\frefformat{vario}{\fancyrefalglabelprefix}{Alg.~#1}

 \newtheorem{thm}{Theorem}
 
 \newtheorem{defi}{Definition}
 \newtheorem{lem}[thm]{Lemma}
 
\safemath{\dictab}{[\,\dicta\,\,\dictb\,]}

\safemath{\ysig}{\bmy}
\safemath{\ysighat}{\hat{\ysig}}
\safemath{\ysigdim}{M}
\safemath{\xsig}{\bmx}
\safemath{\xsigdim}{N}
\safemath{\nx}{n_x}
\safemath{\zsig}{\bmz}
\safemath{\zsigdim}{\ysigdim}
\safemath{\rsig}{\bmr}
\safemath{\Adict}{\bA}
\safemath{\Adicttilde}{\widetilde{\Adict}}
\safemath{\Adictdim}{\outputdim\times\xsigdim}
\safemath{\avec}{\bma}
\safemath{\avectilde}{\tilde{\avec}}
\safemath{\Bdict}{\bB}
\safemath{\Bdicttilde}{\widetilde{\Bdict}}
\safemath{\Cdict}{\bC}
\safemath{\cvec}{\bmc}
\safemath{\Ddict}{\bD}
\safemath{\Ddictdim}{\ysigdim\times\xsigdim}
\safemath{\dvec}{\bmd}
\safemath{\Ddicttilde}{\widetilde{\bD}}
\safemath{\Bonb}{\bB}
\safemath{\bvec}{\bmb}
\safemath{\Bonbdim}{\ysigdim\times\ysigdim}
\safemath{\noise}{\bmn}
\safemath{\noisedim}{\ysigim}
\safemath{\err}{\bme}
\safemath{\errdim}{\ysigdim}
\safemath{\errset}{\setE}
\safemath{\nerr}{n_e}
\safemath{\delop}{\bP_\errset}
\safemath{\delopc}{\bP_{{\errset}^c}}

\safemath{\cplxi}{\imath}
\safemath{\cplxj}{\jmath}

\safemath{\dict}{\matD}
\safemath{\inputdim}{N}		
\safemath{\outputdim}{M}		
\safemath{\sparsity}{S}	
\safemath{\inputdimA}{{N_a}}	
\safemath{\inputdimB}{{N_b}}	
\safemath{\elemA}{{n_a}}	
\safemath{\elemB}{{n_b}}	
\safemath{\resA}{\matR_a}	
\safemath{\resB}{\matR_b}	
\safemath{\subD}{\matS} 
\safemath{\subA}{\matS_a} 
\safemath{\subB}{\matS_b} 
\safemath{\dicta}{\matA} 	
\safemath{\dictb}{\matB} 	
\safemath{\hollowS}{H}
\safemath{\hollowA}{H_a}
\safemath{\hollowB}{H_b}
\safemath{\cross}{Z}
\safemath{\coh}{\mu_d}			
\safemath{\coha}{\mu_a}			
\safemath{\cohb}{\mu_b}			
\safemath{\mubs}{\nu}	
\safemath{\cohm}{\mu_m} 
\safemath{\dictset}{\setD}	
\safemath{\dictsetp}{\dictset(\coh,\coha,\cohb)}	
\safemath{\dictsetgen}{\dictset_\text{gen}}
\safemath{\dictsetgenp}{\dictsetgen(\coh)}
\safemath{\dictsetonb}{\dictset_\text{onb}}
\safemath{\dictsetonbp}{\dictsetonb(\coh)}

\safemath{\leftside}{U}
\safemath{\rightsideA}{R_a}
\safemath{\rightsideB}{R_b}

\safemath{\indexS}{\setI_S} 

\safemath{\na}{n_a}			
\safemath{\nb}{n_b}			
\safemath{\coeffa}{p_i}	
\safemath{\coeffb}{q_j}	
\safemath{\seta}{\setP}		
\safemath{\setb}{\setQ}     
\safemath{\setw}{\setW}	
\safemath{\setz}{\setZ}	
\safemath{\cola}{\veca}		
\safemath{\colb}{\vecb}		
\safemath{\cold}{\vecd}		
\safemath{\inputvec}{\vecx} 	
\safemath{\error}{\vece}	
\safemath{\noiseout}{\vecz} 	
\safemath{\inputvecel}{x}
\safemath{\inputveca}{\vecx_a}
\safemath{\inputvecb}{\vecx_b}
\safemath{\outputvec}{\vecy}	
\safemath{\lambdamin}{\lambda_{\mathrm{min}}}

\safemath{\elltwo}{\ell_2}
\safemath{\ellone}{\ell_1}
\safemath{\ellzero}{\ell_0}
\safemath{\ellinf}{\ell_\infty}
\safemath{\ellinftilde}{\ell_{\widetilde\infty}}
\safemath{\licard}{Z(\coh,\coha,\cohb)}
\safemath{\xsol}{\hat{x}}
\safemath{\xbord}{x_b}		
\safemath{\xstat}{x_s}		
\safemath{\xstatLone}{\tilde{x}_s}
\safemath{\order}{\mathcal{O}} 
\safemath{\scales}{\Theta} 
\safemath{\ones}{\mathbf{1}} 
\safemath{\zeroes}{\mathbf{0}} 
\safemath{\thlone}{\kappa(\coh,\cohb)} 
\safemath{\constoneA}{\delta} 
\safemath{\constoneB}{\epsilon} 
\safemath{\nlarge}{L}				   
\safemath{\sumlarge}{S_\nlarge}
\safemath{\maxlarger}{P_\nlarge}	   
\safemath{\Pzero}{\textrm{P0}}	
\safemath{\Pone}{\textrm{P1}}
\safemath{\vecfir}{\vecw}			 
\safemath{\vecsec}{\vecz}
\safemath{\elvecfir}{w}              
\safemath{\elvecsec}{z}				 
\safemath{\nlargefir}{n}
\safemath{\normout}{\gamma}
\safemath{\auxfun}{h}
\safemath{\supp}{\textrm{supp}}

\safemath{\indexa}{\ell}
\safemath{\indexb}{r}
\safemath{\indexc}{i}
\safemath{\indexd}{j}

\safemath{\project}{P}

\newcommand*{\fancyrefremarklabelprefix}{remark}
\frefformat{vario}{\fancyrefremarklabelprefix}{Remark~#1}

\IEEEoverridecommandlockouts 
\allowdisplaybreaks 
\usepackage{color}

\newcommand{\proj}[2]{\mathrm{proj}_{#1}\!\left(#2\right)}

\usepackage{bbm}

\def\setIc{{\mathcal{I}^c}}
\def\PINC{\textit{PINC}}

\makeatletter

\newcommand{\leqnomode}{\tagsleft@true}
\newcommand{\reqnomode}{\tagsleft@false}

\makeatother

\title{Alternating Projections Method for Joint Precoding \\ and Peak-to-Average-Power Ratio Reduction}
\author{\IEEEauthorblockN{Sueda Taner and Christoph Studer} \\[0.05cm]
\thanks{The work of ST and CS was supported in part by ComSenTer, one of six centers in JUMP, an SRC program sponsored by DARPA. The work of CS was also supported in part by an ETH Research Grant and by the US National Science Foundation (NSF) under grants CNS-1717559 and ECCS-1824379.}
\thanks{The authors would like to thank Artur Gorokh for discussions on the projection onto the PAR-bounded set and Gian Marti for comments on an early version of the paper.}
\IEEEauthorblockA{\em Department of Information Technology and Electrical Engineering, ETH Zurich, Switzerland \\
e-mail: taners@iis.ee.ethz.ch and studer@ethz.ch} 
}

\begin{document}

\maketitle


\begin{abstract}

Orthogonal frequency-division multiplexing (OFDM) time-domain signals exhibit high peak-to-average (power) ratio (PAR), which requires linear radio-frequency chains to avoid an increase in error-vector magnitude (EVM) and out-of-band (OOB) emissions. 
In this paper, we propose a novel joint PAR reduction and precoding algorithm that relaxes these linearity requirements in massive multiuser (MU) multiple-input multiple-output (MIMO) wireless systems.
Concretely, we develop a novel alternating projections method, which limits the PAR and transmit power increase while simultaneously suppressing MU interference.
We provide a theoretical foundation of our algorithm and provide simulation results for a massive MU-MIMO-OFDM scenario. Our results demonstrate significant PAR reduction while limiting the transmit power, without causing EVM or OOB emissions.  
\end{abstract}



\section{Introduction} 
\label{sec:intro}

Massive multiuser (MU) multiple-input multiple-output (MIMO) is a popular technology to increase spectral efficiency~\cite{larsson14a}. Orthogonal frequency-division multiplexing (OFDM)~\cite{NP00} is widely used to deal with inter-symbol-interference.
While the combination of massive MU-MIMO with OFDM has the potential to achieve high spectral efficiency, OFDM time-domain signals  exhibit  high dynamic range~\cite{HL05}, which requires \emph{linear} radio-frequency (RF) chains to avoid an increase in error-vector magnitude (EVM) and out-of-band (OOB) emissions.
Unfortunately, such systems would require a large number of linear RF chains, which quickly results in excessively high RF circuit power consumption and system costs. 
To address this issue, a plethora of peak-to-average (power) ratio (PAR) reduction methods have been proposed in the OFDM literature; see, e.g.,~\cite{rahmatallah_survey} and the references therein. 

Among the recent literature, the methods in~\cite{bao18,kant2021asilomar,kant2022twc,gokceli2022,kalinov21,jang2021globecom} modify the transmitted OFDM time-domain signals to reduce the PAR.
Alternatively, as shown in~\cite{studer13a}, the excess degrees-of-freedom of  massive MU-MIMO systems enables the design of precoding algorithms at the basestation (BS) that jointly reduce the PAR in the transmitted signals while canceling MU interference (MUI). 
To reduce the complexity of such emerging joint precoding and PAR reduction (JPP) methods, a number of algorithms have been proposed~\cite{cha14,wang14,studer15b,guo16,zayani19,bao16,yao19,liu19,hua2022ladmm}.
However, such methods either increase MUI, EVM, or OOB emissions, or aim at minimizing the transmit signal's peaks (e.g., their $\ell^\infty$-norm), but not the actual PAR. 
Only recently, a novel $\ell^p\!-\!\ell^q$-norm minimization approach was proposed in~\cite{taner2021ellpellqnorm}, which is capable of finding minimum-PAR solutions unachievable by $\ell^\infty$-norm minimization methods without {causing} EVM, assuming perfect transmit-side channel-state information (CSI), or OOB emissions.
Reference~\cite{taner2021ellpellqnorm} also shows the existence of a fundamental trade-off between the PAR and the transmit power of the OFDM time-domain signals, so that reducing the PAR must come at the cost of a power increase (PINC), assuming that the EVM should remain zero.
For a given transmit power constraint, this means that one has to back-off in transmit power, which lowers the signal-to-noise ratio (SNR) at the user-equipment (UE) side. Unfortunately, the $\ell^p\!-\!\ell^q$-norm minimization approach from~\cite{taner2021ellpellqnorm} provides no control over the PINC of the transmitted time-domain signals.

\subsection{Contributions}
We propose a novel JPP method for massive MU-MIMO-OFDM systems based on an alternating projections method (APM) that enables precise control over the PAR and PINC.
Our proposed algorithm is not only capable of finding solutions with lower PAR than the widely-used  $\ell^\infty$-norm minimization approaches in~\cite{cha14,wang14,studer15b,guo16,zayani19,bao16,yao19,liu19,bao18}, but also enables lower PINC than the recent $\ell^p\!-\!\ell^q$-norm minimization approach from~\cite{taner2021ellpellqnorm}.
We demonstrate the efficacy of our approach for a massive MU-MIMO-OFDM scenario by showing that our APM achieves low PAR and low PINC, while perfectly removing MUI without causing EVM (assuming perfect CSI at the transmitter) or OOB emissions compared to least-squares precoding.

\subsection{Notation}
We represent column vectors and matrices by bold lowercase and uppercase letters, respectively.
The $k$th entry of a vector~$\veca$ is~$a_k$ and the $k$th column of a matrix $\bA$ is $[\bA]_k=\veca_k$.
The matrix transpose and Hermitian are designated by the superscripts $(\cdot)^T$ and $(\cdot)^H$, respectively.
We use $\mathbf{0}_{N\times M}$ for the $N\times M$ all-zeros matrix and $\bF_N$ for the  $N\times N$ unitary discrete Fourier transform (DFT) matrix.
We define the $\ell^p$-norm as $\vecnorm{\bma}_p=(\sum_{k} |a_k|^p )^{1/p}$ and the Frobenius norm as $\|\bA\|_F=\sqrt{\sum_k \|\veca_k\|_2^2}$. 
We use calligraphic letters to denote sets and the superscript $(\cdot)^c$ for the set complement.
Given a subset of indices $\setI$, $\veca_\setI$ represents the vector whose entries are given by $\{a_i\}_{i\in\setI}$.
The orthogonal projection of a vector $\veca$ onto a set $\setA$ is denoted by $\proj{\setA}{\veca}$.
%


\section{Prerequisites}
\label{sec:prereq}

The joint precoding and PAR reduction problem in MU-MIMO-OFDM is a special case of a PAR reduction problem with linear constraints.
Thus, we first review the PAR minimization problem under a linear constraint $\bmy=\bA\bmx$ and study the limits of existing algorithms that find solution vectors~$\bmx$ with low (or minimal) PAR in Sections \ref{sec:prereq} and \ref{sec:problem}. We then apply the problem setup to the special, but more complicated case of an MU-MIMO-OFDM scenario in \fref{sec:ofdm}.
 
\subsection{Peak-to-Average (Power) Ratio (PAR)}

Given an underdetermined system of linear equations $\bmy=\bA\bmx$, where $\bmy\in\complexset^M$ and $\bA\in\complexset^{M\times N}$ is full-rank with $M< N$, we wish to compute solution vectors $\vecx\in\opC^N$ with low dynamic range. 
In communication applications, the dynamic range is commonly measured  by the peak-to-average-power ratio (PAR)
\begin{align} \label{eq:defPAR}
\textit{PAR}(\bmx) \define \frac{N\|\bmx\|^2_\infty}{\|\bmx\|^2_2}.
\end{align}
for non-zero vectors $\bmx\in\complexset^N$.
The PAR satisfies $1 \leq \textit{PAR}(\bmx) \leq N$, where the lower bound holds with equality if and only if $|x_i|=|x_j|$ for all $i,j\in\{1,\ldots,N\}$. We will call such $\vecx$ a ``minimum-PAR (min-PAR) vector'' throughout this paper.

\subsection{Computing Low-PAR Solutions via $\ell^\infty$-Norm Minimization}
Since the PAR in \fref{eq:defPAR} is nonconvex and nondifferentiable, directly minimizing $\PAR(\vecx)$ subject to the affine constraint $\bmy=\bA\bmx$ is difficult.
Thus, references~\cite{studer13a,studer15b} proposed to minimize the $\ell^\infty$-norm of solution vectors:
\leqnomode
\begin{align} \label{eq:linftyminimization}
\qquad\quad\,\,\,\hat\bmx^\infty = \argmin_{\tilde\bmx\in\complexset^N}\, \|\tilde\bmx\|_\infty \quad \text{subject to } \bmy=\bA\tilde\bmx. \tag{P-$\infty$}
\end{align}
\reqnomode
Although $\ell^\infty$-norm minimization provably reduces the PAR~\cite{studer15b}, such methods only reduce the signal's peaks and do, in general, not find min-PAR solutions. 
To compute min-PAR solutions, an $\ell^p\!-\!\ell^q$-norm problem formulation was proposed recently in~\cite{taner2021ellpellqnorm}. 
In \fref{sec:problem}, we propose a novel method that outperforms such approaches by directly dealing with the following trade-off between PAR and PINC.

\subsection{Fundamental PAR vs.\ PINC Trade-off}
\label{sec:parpinctradeoff}

Due to the power constraints of transmitters, we must also consider the power of the solution vectors along with their PAR.
The least-squares (LS) vector $\hat\vecx^{\text{LS}} \triangleq \bA^H(\bA\bA^H)^{-1}\vecy$, has the minimum power $\|\bmx\|_2^2$, among all the solutions $\bmx$ to $\vecy=\bA\vecx$, by definition.
Other solution vectors~$\bmx$, e.g., vectors with low PAR which suit nonlinear RF circuitry better, typically have higher power than~$\hat\vecx^{\text{LS}}$. 
In~\cite{taner2021ellpellqnorm}, this observation was made explicit by a fundamental trade-off between the PAR of a solution vector~$\bmx$ satisfying $\bmy=\bA\bmx$ and its power compared to the LS solution using the following definition~\cite{taner2021ellpellqnorm}:
\begin{defi}
Let $\bmx$ be any solution vector to $\bmy=\bA\bmx$ and $\hat\bmx^\text{LS} = \bA^H(\bA\bA^H)^{-1}\vecy$. Then, the power increase (PINC) of the vector $\bmx$ is defined as $\textit{PINC}(\bmx) \define {\|\bmx\|_2^2}/{\|\hat\bmx^\text{LS}\|_2^2}.$
\end{defi}
The following result, taken from~\cite[Lem.~1]{taner2021ellpellqnorm}, reveals this fundamental trade-off between PAR and PINC.
\begin{lem} \label{lem:PARPINCtradeoff}
Fix $\bA$ and $\bmy$, and let $\bmx$ be any nonzero solution vector to $\bmy=\bA\bmx$. Then, there exists a constant $c\geq1$ (for fixed~$\bmy$ and~$\bA$) which satisfies the following inequality: 
\begin{align} \label{eq:tradeoffomg}
\textit{PAR}(\bmx) \, \textit{PINC}(\bmx) \geq c.
\end{align}
\end{lem}

Since the solution vector $\hat\bmx^\infty$ from~\fref{eq:linftyminimization} achieves the lower bound in \fref{eq:tradeoffomg} with equality and  defines the constant  $c= {\|\hat\vecx^\infty\|_\infty^2}/{\|\hat\vecx^{\text{LS}}\|_2^2}$ \cite{taner2021ellpellqnorm}, the vector $\hat\bmx^\infty$ is optimal in the PAR-PINC trade-off, but would not generally have minimal PAR. Furthermore, min-PAR solutions typically have higher PINC than other solutions.
Hence, to normalize the transmit power according to a given power constraint, the transmitter must back-off more compared to the power of the LS solution, which will lower the SNR at the UE side. 

Although the $\ell^p\!-\!\ell^q$-norm minimization approach in~\cite{taner2021ellpellqnorm} was shown to find min-PAR solutions, it provides no control over the PINC of the solutions.
We next design a novel method that enables precise control over both the PAR and PINC.


\section{PAR and PINC Reduction with an APM}
\label{sec:problem}

We now introduce our approach {to find solution vectors~$\bmx$} with low PAR \emph{and} low PINC for the general case of  $\vecy = \bA\vecx$. A concrete application of our approach to JPP in a massive MU-MIMO-OFDM scenario is shown in \fref{sec:ofdm}.

\subsection{Computing Solutions with  Bounded PAR and PINC}
\label{sec:general_a}
Our approach builds upon the following feasibility problem to find solution vectors  $\vecx$ that (i) satisfy $\vecy=\bA\vecx$, (ii) have bounded PAR, and (iii) have bounded PINC:
\begin{align*} 
(\text{P-F}) \,\,
 \left\{\begin{array}{ll} 
\text{find} &  \vecx \in \complexset^N\\
\text{subject to} & \vecy = \bA\vecx, \, \PAR(\vecx) \leq \rho, \, \PINC(\vecx)\leq \xi.
\end{array}\right.
\end{align*}
\reqnomode
Here, the parameters $\rho$ and $\xi$ denote the desired PAR and PINC bounds satisfying $1\leq\rho\leq N$ and $1\leq \xi$, respectively.

Given the sets $\setC,\setD\subseteq \opC^N$, APMs typically solve constrained feasibility problems of the form~\cite{ginat2018method} 
\begin{align}
\text{find }\vecx\in\opC^N \text{ subject to } \bmx\in \setC\cap \setD,
\end{align}
by alternating between the projection onto sets $\setC$ and $\setD$ iteratively for iterations $k=1,2,\ldots$ until convergence: 
\begin{align} \label{eq:apm}
\vecx^{(k)} = \proj{\setC}{\proj{\setD}{\vecx^{(k-1)}}\!}\!.
\end{align}

In order to solve (P-F),  we define the following sets: 
\begin{align}
\setC&\triangleq \{\vecx\in\opC^N \, |\, \vecy=\bA\vecx\},  \label{eq:setC}\\
\setD&\triangleq\{\vecx\in\opC^N \,|\,\PAR(\vecx)\leq\rho, \PINC(\vecx)\leq \xi\}. \label{eq:setD} 
 \end{align}
 We detail the projection operators required in \fref{eq:apm} below.
To calculate solutions $\bmx$ with bounded PAR and PINC, we run our APM for a fixed number of iterations~$K_{\text{max}}$. 
{We emphasize that $\vecy=\bA\vecx^{(k)}$ is always satisfied since $\proj{\setC}{\cdot}$ is carried out \emph{after} evaluating $\proj{\setD}{\cdot}$ in each iteration.
For the same reason, $\vecx^{(k)}$ does not necessarily satisfy the PAR-PINC constraints defined by $\setD$, and our APM procedure computes only approximate solutions to (P-F) as there are no guarantees on feasibility or convergence due to nonconvexity of the optimization problem. 

\subsection{The Orthogonal Projection Operators}\label{sec:apm}

\subsubsection{Projection onto the Set $\setC$} 

\label{sec:projC}
The orthogonal projection of a vector $\vecz\in\opC^N$ onto the set $\setC$  defined in \fref{eq:setC}  is given by \cite{GSB14}:
\begin{align} \label{eq:proximal}
\proj{\setC}{\vecz}  &= \bmz - \bA^H(\bA\bA^H)^{-1}(\bA\vecz-\vecy).
\end{align}
Note that $\proj{\setC}{\bZero_{N\times 1}}=\hat\bmx^\text{LS}$. Hence, we initialize $\vecx^{(0)}=\bZero_N$ and, only in the first APM iteration,  skip $\proj{\setD}{\vecx^{(0)}}$  to ensure that our method starts from the LS solution ${\vecx}^{(1)}=\hat\bmx^\text{LS}$.

\subsubsection{Projection onto the Set $\setD$}
\label{sec:projD}

Using $\hat\bmx^\text{LS}$ from the first APM iteration, we compute $\vecx = \proj{\setD}{\vecz}\in\opC^N$ with the steps listed below. 
A rigorous derivation of the following projection algorithm can be found in \fref{app:KKTproof}.
\begin{enumerate}[\arabic*:] 
\item Set $P= \xi\|\hat\bmx^\text{LS}\|_2^2$.  
\item If $\PAR(\vecz)\leq \rho $, then {set $\vecx' = \vecz$ and $P'=\|\vecx'\|_2^2$, and skip to Step 5.}
Otherwise, set $\alpha = \rho/N$ and initialize $L=1$.
\item Let the set $\setI$ index the $L$ entries of $\vecz$ with the largest magnitude. If (i) this set is not uniquely determined, or 
(ii) $\max_{i\in\setIc}|z_i| \leq \sqrt{\frac{\alpha}{1-\alpha L}}\|\vecz_\setIc\|_2 < \min_{i\in\setI}|z_i| $ is not satisfied, then increment $L$ by 1 and repeat this step. 
\item If $\vecz_\setIc=\bZero_{(N-L)\times 1}$, then compute $\vecx' \in\opC^N$ using
\begin{align}
x_i' = \begin{cases} 
\sqrt{\frac{(1-\alpha L)P'}{N-L}}, \, &i \in \setIc\\
\frac{\sqrt{\alpha P'} }{|z_i|} z_i, \, &i \in \setI,
\end{cases}
\end{align}
where {$P' = \alpha \|\vecz_\setI\|_1^2$. }
Otherwise, compute $\vecx'\in\opC^N$ using
\begin{align}
x_i' = \begin{cases} 
\frac{\sqrt{(1-\alpha L)P'}}{\|\vecz_\setIc\|_2}z_i, \, &i \in \setIc\\
\frac{\sqrt{\alpha P'} }{|z_i|} z_i, \, &i \in \setI,
\end{cases}
\end{align}
where {$P' = \big( \sqrt{1-\alpha L}\|\vecz_\setIc\|_2 + \sqrt{\alpha}\|\vecz_\setI\|_1\big)^2$.}

\item {$ \vecx=\min\!\big\{1,\sqrt{{P}/{P'}}\big\}\vecx'$.}

\end{enumerate}
Since Step~3 requires sorting of the magnitudes of the entries of~$\vecz$, the complexity of computing $\proj{\setD}{\vecz}$ is  $O(N\log(N))$.
We conclude by noting that a special case of this procedure was given in~\cite[Alg.~2]{tropp2005}, where an equality constraint is imposed on $\|\vecx\|_2^2$ instead of an upper bound---this reduces the PAR constraint to an $\ell^\infty$-norm constraint, which is different from $\proj{\setD}{\cdot}$ and easier to solve. However, 
imposing an upper bound instead of an equality has the advantage of computing solutions with lower power, if they exist, in congruence with our aim of keeping the PINC small.

\subsection{Example of PAR Reduction with our APM}
\label{sec:simpleex}

In \fref{fig:simplecase}, we show an example of PAR reduction with our APM. We apply our algorithm to one fixed instance of a circularly-symmetric complex standard normal matrix $\bA\in\opC^{100\times 200}$ and vector $\vecy\in\opC^{100}$.
We consider the cases where the PAR and PINC bounds are: (i) $\rho_{\text{dB}}=0.4$ and $\xi_{\text{dB}}=1.6$, and (ii) $\rho_{\text{dB}}=0.2$ and $\xi_{\text{dB}}=2$ in decibel. 
As a baseline, we also show $\ell^\infty$-norm minimization via CRAMP~\cite{studer15b}\footnote{We select one $\ell^\infty$-norm minimization algorithm out of the many in the literature to demonstrate the limitation of such methods for reducing the PAR.} and $\ell^2\!-\!\ell^1$-norm minimization from~\cite{taner2021ellpellqnorm}.
All algorithms are initialized with $\bmx^{(1)}=\proj{\setC}{\bZero_{100\times 1}}=\hat\bmx^\text{LS}$ and run until convergence. 

\fref{fig:simplecase} shows the PAR vs.~PINC trade-off for all iterations together with the lower-bound given by \fref{lem:PARPINCtradeoff}.
We observe that the $\ell^\infty$-norm minimization approach converges to a solution whose PAR is 0.7\,dB, and has smaller PINC than the other methods for this PAR value. However, the min-PAR solutions, i.e., solutions whose PAR is 0\,dB, are attained by our APM and $\ell^p\!-\!\ell^q$-norm minimization---but, as expected, at the cost of higher PINC.  
This example effectively demonstrates that the APM and $\ell^p\!-\!\ell^q$-norm minimization methods are more suitable to \textit{minimize} the PAR than $\ell^\infty$-norm minimization, which instead yields solutions achieving the optimal PAR-PINC tradeoff. 
We also observe that the proposed APM outperforms  $\ell^p\!-\!\ell^q$-norm minimization in the PAR-PINC trade-off: APM with $\rho_{\text{dB}}=0.4,\xi_{\text{dB}}=1.6$ and $\rho_{\text{dB}}=0.2,\xi_{\text{dB}}=2$ converge to  0.4\,dB and 0.2\,dB PAR with 1.4\,dB and 1.2\,dB lower PINC than $\ell^2\!-\!\ell^1$-norm minimization, respectively. 
This example showcases the APM's advantage over $\ell^p\!-\!\ell^q$-norm minimization in causing lower PINC while reducing the PAR of solutions.

\begin{figure}[tp]
\centering
\includegraphics[width=0.85\columnwidth]{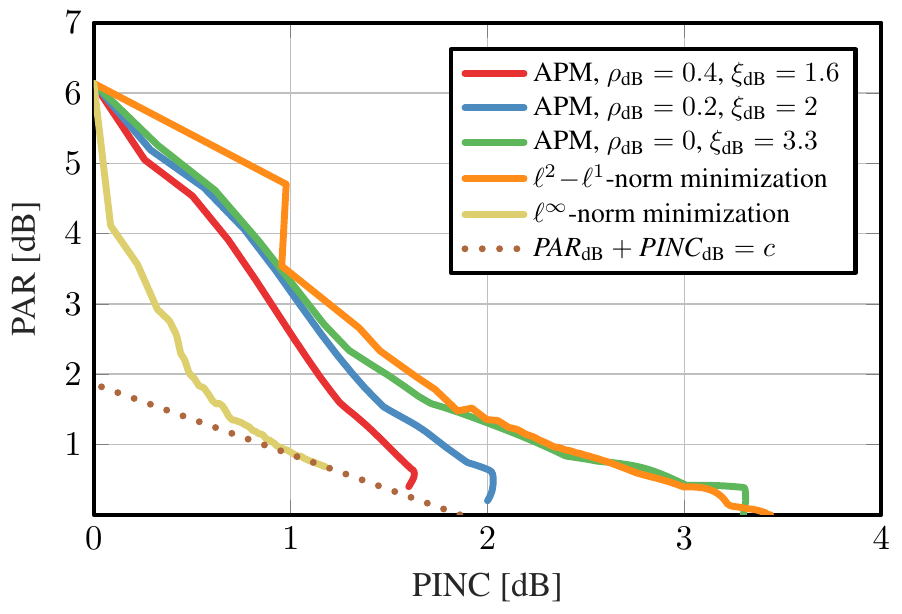}\label{fig:simpletradeoff}\\[-0.2cm]
\caption{Trade-off between PAR and PINC for the proposed APM, $\ell^p\!-\!\ell^q$-norm minimization~\cite{taner2021ellpellqnorm}, and $\ell^\infty$-norm minimization solved via CRAMP~\cite{studer15b}. The APM achieves the same PAR as $\ell^2\!-\!\ell^1$-norm minimization at lower PINC.}
 \label{fig:simplecase}
\end{figure}

\section{The Massive MU-MIMO-OFDM Case}
\label{sec:ofdm}

We now show an application example of the proposed APM for JPP in a massive MU-MIMO-OFDM scenario. 

\subsection{System Model}
\label{sec:sys_model}
We consider the downlink of a massive MU-MIMO-OFDM system where a $B$-antenna BS transmits data to~{$U$ single-antenna UEs with $U<B$}.
We assume that the total number of OFDM subcarriers is $W,$ and $\setW$
and $\setW^c$ designate the sets of used and unused OFDM subcarriers, respectively, where $|\setW|+|\setW^c|=W$.
For the used subcarriers, the signal vectors $\vecs_w\in\mathcal{S}^U$, $w\in\setW$, contain the symbols to be transmitted from the constellation $\mathcal{S}$ for each UE.
For the unused subcarriers, we set $\vecs_w=\mathbf{0}_{U\times 1}$, $w\in\setW^c$.
The signal vectors $\vecs_w$, $w\in\setW$, are precoded into $W$ frequency-domain vectors $\vecx_w\in\opC^B$ with the goal of suppressing MUI.
We define the matrix $\bX\in\opC^{B\times W}$ as $\bX\define [\bmx_1,\ldots,\bmx_W]$.
The frequency-domain outputs of each BS antenna are given by $\bX^T$, whose columns correspond to BS antennas and rows to OFDM subcarriers.
The total transmit power $\|\bX\|_F^2$ depends on the transmit signals $\vecs_w$, $\forall w$, and the channel state. Hence, we assume that the precoded vectors will be normalized prior to transmission via $\hat\vecx_w = \vecx_w /\|\bX\|_F$ to ensure unit transmit power.
While this normalization step is necessary to satisfy regulatory power constraints, we omit this step while describing our precoder and recall it in~\fref{sec:results}.

Let $\vect_b$ denote the time-domain output of the $b$th BS antenna and define the matrix $\bT \in\opC^{W\times B}$ by $\bT\define [\bmt_1,\ldots,\bmt_B]$.
In an OFDM system, the matrix $\bT$ is given by the inverse DFT as $\bT = \bF^H\bX^T$.
To simplify notation, we define the linear mapping from $\{\vect_b\}_{b=1}^B$ to $\{\vecx_w\}_{w=1}^{W}$ as follows:
\begin{align} \label{eq:psithemapper}
\psi_w(\vect_1,\dots,\vect_B) \define  [(\bF\bT)^T]_w=\vecx_w,\, w\in\{1,\dots,W\}.
\end{align}
In order to avoid ISI, a cyclic prefix  is prepended to each time-domain vector $\vect_b$, $b\in \{1,\dots,B\}$, prior to transmission.

Let $\vecy_w\in\opC^W$ denote the received vector at OFDM subcarrier $w$. We model $\vecy_w$ with the following input-output relation of the wireless channel in the frequency domain: 
\begin{align} \label{eq:y_Hs}
    \vecy_w = \bH_w \vecx_w + \vecn_w,\, w\in\{1,\dots,W\}.
\end{align}
Here, $\bH_w\in\opC^{U\times B}$ denotes the MIMO channel matrix associated with the $w$th subcarrier and $\vecn_w\in\opC^U$ models circularly-symmetric Gaussian noise.
Finally, OFDM demodulation is performed by each of the $U$ UEs in order to estimate the transmitted data symbols, i.e., data detection is carried out using $[\vecy_w]_u$ for UE $u$ at each used subcarrier $w \in\setW$.  

\subsection{Least-Squares Precoding}
\label{sec:LSprecoding}
The BS must employ precoding in order to suppress MUI.
Linear precoding simply computes $\vecx_w = \bG_w \vecs_w$ on the used subcarriers $w\in\setW$ with the precoding matrix $\bG_w\in\opC^{B\times U}$, and set $\bmx_w=\bZero_{B\times 1}$ on the unused subcarriers $w\in\setW^c$.
{LS precoding is widely used and minimizes the transmit power while satisfying the following precoding constraints (under the assumption that the channel matrices $\bH_w$, $\forall w$, are known perfectly at the BS-side}\footnote{We assume perfect channel knowledge for simplicity. However, one can also use an estimate of the channel acquired by exploiting reciprocity. In the latter scenario, note that there would be nonzero EVM due to channel estimation errors, however, the EVM \textit{caused by} precoding would still be zero.})
(PC1) $\vecs_w = \bH_w\vecx_w$, $w\in\setW$, which ensure zero EVM, and (PC2) $\bmx_w=\bZero_{B\times 1}$, $w\in\setW^c$, which ensure zero OOB emissions. 
For the LS precoder, the precoding constraints in (PC1) have the  closed-form solution $\vecx_w^{\text{LS}} = \bG_w^{\text{LS}} \vecs_w$ with $\bG_w^{\text{LS}}=\bH_w^H (\bH_w\bH_w^H)^{-1}$, $w\in\setW$.
Although LS precoding perfectly eliminates MUI and minimizes the transmit power, with $\textit{PINC}=1$ by definition, the PAR of the resulting time-domain signals is typically high~\cite{studer13a,cha14,wang14,studer15b,guo16,zayani19,bao16,yao19,liu19,bao18}.
We next utilize the proposed APM to mitigate this drawback.

\subsection{Joint Precoding and PAR Reduction with our APM}
\label{sec:precoding}

Massive MU-MIMO systems have the unique property that the downlink channel matrices have a large nullspace, which can be exploited to simultaneously satisfy the precoding constraints (PC1) and (PC2), while shaping the transmitted time-domain signals to reduce the PAR~\cite{studer13a}.
Our goal is to find frequency-domain precoding vectors $\vecx_w$ such that the time-domain signals are bounded in both PAR and PINC, while the vectors~$\vecx_w$ satisfy (PC1) and (PC2).
To achieve all of these goals, we propose the following feasibility problem:
\begin{align*} 
(\text{JPP-F}) \,\,\left\{\begin{array}{ll} 
\text{find} & \displaystyle \vect_1,\dots,\vect_B \in \complexset^W \\[0.1cm]
\text{subject to} & \vecs_w = \bH_w\psi_w(\vect_1,\dots,\vect_b) ,  w\in\setW\\[0.1cm]
& \mathbf{0}_{U\times 1} =\psi_w(\vect_1,\dots,\vect_b) ,\,  w\in\setW^c\\[0.1cm]
& \PAR(\vect_b)\leq \rho,\,b\in \{1,\dots,B\}\\[0.1cm]
& \PINC(\bT) \leq \xi.
\end{array}\right.
\end{align*}
Here, we separately minimize the PAR at each transmit antenna and consider the total power of the transmit signals compared to the LS solution by defining $\PINC(\bT) \define \|\bT\|_F^2/\|\bT^{\text{LS}}\|_F^2$.

The remaining piece of the puzzle is to find a solution to (JPP-F) efficiently (and approximately) via  our APM. Our approach is merely a more complicated version of the APM proposed in \fref{sec:problem} and the details are as follows. 
Since there exists a one-to-one mapping between time and frequency via~\fref{eq:psithemapper},
we can apply the linear constraints in (JPP-F)  separately to the columns of the frequency domain matrix $\bX$. Here, we initialize with $\vecx_w^{(0)}=\bZero_{B\times 1},\forall w$, which calculates the LS solution in the first iteration, i.e., $\bX^{(1)}=\bX^{\text{LS}}$. 
At every iteration, we use~\fref{eq:proximal} to compute $\vecx_w^{(k)}$ for $w\in\setW$ and set $\vecx_w^{(k)}=\mathbf{0}_{U\times 1}$ for $w\in\setW^c$. 
As in~\fref{sec:apm}, we utilize $\|\bX^{\text{LS}}\|_F^2=\|\bT^{\text{LS}}\|_F^2$ to rewrite the PINC constraint of (JPP-F) as a power constraint on $\|\bT\|_F^2$.  
We project the time-domain vectors onto the PAR-bounded set individually for each antenna and then scale $\bT^{(k)}$ to satisfy the PINC constraint. 
{Analogous to our algorithm for the general case in \fref{sec:general_a}, we take the projection onto the linear (precoding) constraints as the output of each iteration, so that (PC1) and (PC2) are always satisfied. This implies that the APM does not increase the EVM or OOB emissions compared to LS precoding, while decreasing the PAR.}
Note that our APM calculates \textit{approximate} solutions to (JPP-F) with no guarantees on feasibility or convergence due to its nonconvex nature.
 

\section{Simulation Results}
\label{sec:results}

We now demonstrate the efficacy of the proposed APM for JPP in a massive MU-MIMO-OFDM system. 
\subsection{Simulation Setup}
\label{sec:simulsetup}
As in~\cite{taner2021ellpellqnorm}, we consider a massive MU-MIMO-OFDM system with $B=128$ BS antennas, $U=16$ single-antenna UEs, and 16-QAM transmission. 
We consider  $W=2048$ subcarriers and a bandwidth of 20\,MHz; the used and unused subcarriers are defined in \cite{3gpp19a}.
We assume a simple channel model with $4$ taps, where the entries of the non-zero time-domain matrices are assumed i.i.d.\ circularly complex Gaussian with unit variance. The frequency-domain channel matrices are obtained using the Fourier transform~\cite{studer16a}.
As mentioned in \fref{sec:sys_model}, the precoded vectors will be normalized to unit power prior to transmission.
This back-off in transmit power is equivalent to an $\textit{SNR}_\text{dB}$ loss of exactly $\textit{PINC}_\text{dB}$ compared to the LS precoder.

We solve (JPP-F) using our APM as explained in \fref{sec:precoding} and consider two parameter settings (i) $\rho_{\text{dB}}=3$, $\xi_{\text{dB}}=0.3$ and (ii) $\rho_{\text{dB}}=4$, $\xi_{\text{dB}}=0.1$. 
As baselines, we also compare with $\ell^\infty$-norm minimization solved using  CRAMP~\cite{studer15b} and $\ell^4\!-\!\ell^2$-norm minimization\footnote{We use the $\ell^4\!-\!\ell^2$-norm  instead of the $\ell^2\!-\!\ell^1$-norm, as it was shown in~\cite{taner2021ellpellqnorm} to outperform the $\ell^2\!-\!\ell^1$-norm-based approach in this scenario.} from~\cite{taner2021ellpellqnorm}.
All algorithms run for  $K_\text{max}=20$ iterations.
We reiterate that none of these algorithms increase the EVM or OOB emissions compared to LS precoding; this implies that in the resulting error-rate performance, the SNR loss compared to the LS precoder is determined solely by the PINC. 
Concretely, these JPP algorithms require exactly $\textit{PINC}_\text{dB}$ higher SNR than LS precoder to achieve the same BER.

\subsection{Results and Discussion}

\fref{fig:pinc_par_tradeoff} shows the trade-off between PINC and PAR for the JPP methods and the massive MU-MIMO-OFDM scenario described in~\fref{sec:simulsetup}.
We use Monte-Carlo sampling to compute the complementary cumulative distribution function (CCDF) for the PAR at each antenna's time domain output and the total PINC at each iteration for the different JPP methods.
The CCDF of a random variable $Z$ is defined as $\text{CCDF}_Z(z) = \mathbb{P}(Z > z)$. For example, the value $z$ for which $\text{CCDF}_Z(z) = 1$\% is the 99th percentile of~$Z$. 
In order to demonstrate the per-iteration \textit{pessimistic} behavior of the JPP algorithms, we pick the 99th percentile as the operating point for both $\textit{PAR}_{\text{dB}}$ and $\textit{PINC}_{\text{dB}}$, and show the PAR-PINC trade-off for each algorithm iteration in \fref{fig:pinc_par_tradeoff}, where the iterations are indicated by markers.
Here, we {no longer have the notion of an optimal trade-off since the vectors whose PAR we aim to reduce are not directly the solutions of one linear system (as done in the example of \fref{sec:simpleex}). We therefore plot the 99th percentile of PAR and PINC over $100$ randomized trials}. 
Analogously to \fref{sec:simpleex}, we observe that  our APM and the $\ell^4\!-\!\ell^2$-norm method are able to compute lower-PAR solutions than $\ell^\infty$-norm minimization. 
Since we perform only 20 iterations, we do not expect the APM variants to satisfy the specified PAR and PINC bounds $\rho_{\text{dB}}$ and $\xi_{\text{dB}}$, respectively. 
Nonetheless, the proposed APM has the advantage of achieving $4.7$\,dB PAR at $1$\,dB and $1.1$\,dB lower PINC than $\ell^\infty$-norm minimization, with $\rho_{\text{dB}}=3,\xi_{\text{dB}}=0.3$ and $\rho=4,\xi_{\text{dB}}=0.1$, respectively;
{this demonstrates our APM's capability to find solutions with lower PINC than $\ell^p\!-\!\ell^q$-norm minimization at the same PAR.}

{\fref{fig:ccdf_iter5} shows the CCDF of PAR and PINC for the $5$th iteration 
of the considered JPP algorithms together with the CCDF resulting from LS precoding. 
In only five iterations, all of the iterative JPP methods decrease the PAR at a CCDF target value of 1\% by at least 5\,dB compared to LS precoding, while the APM variants result in at least 0.4\,dB lower PINC than $\ell^\infty$-norm and  $\ell^4\!-\!\ell^2$-norm minimizing approaches.
Thus, our APM is a promising low-complexity PAR reduction method for massive MU-MIMO-OFDM systems with power constraints.

\begin{figure}[tp]
\centering
\includegraphics[width=0.85\columnwidth]{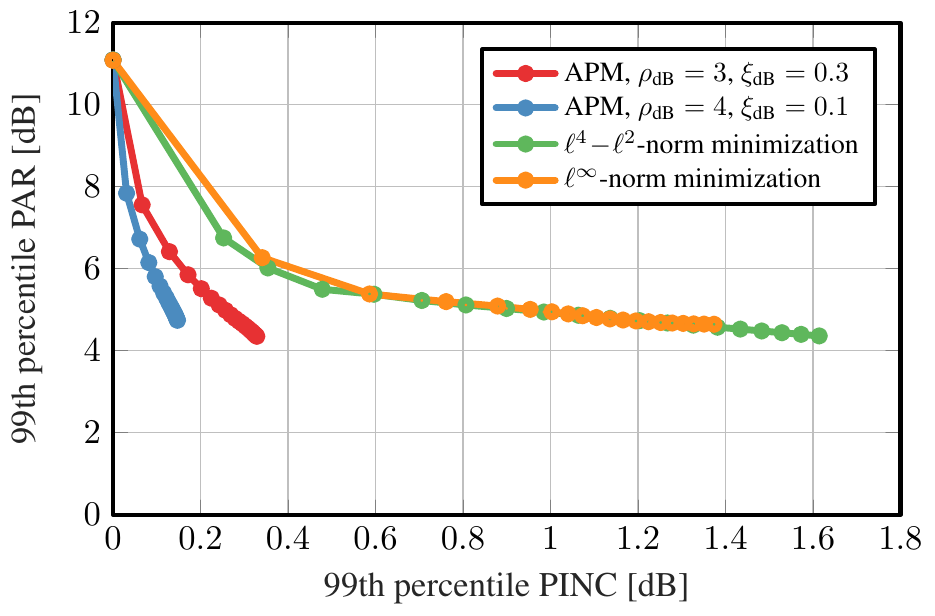}
\vspace{-0.2cm}
\caption{PAR vs.\ PINC trade-off for PPB-based JPP methods in a massive MU-MIMO-OFDM system. The markers correspond to iterations and all algorithms start with the LS solution at $11.1$\,dB PAR and $0$\,dB PINC.}
\label{fig:pinc_par_tradeoff}
\end{figure}

\begin{figure}[tp]
\centering
\includegraphics[width=0.95\columnwidth]{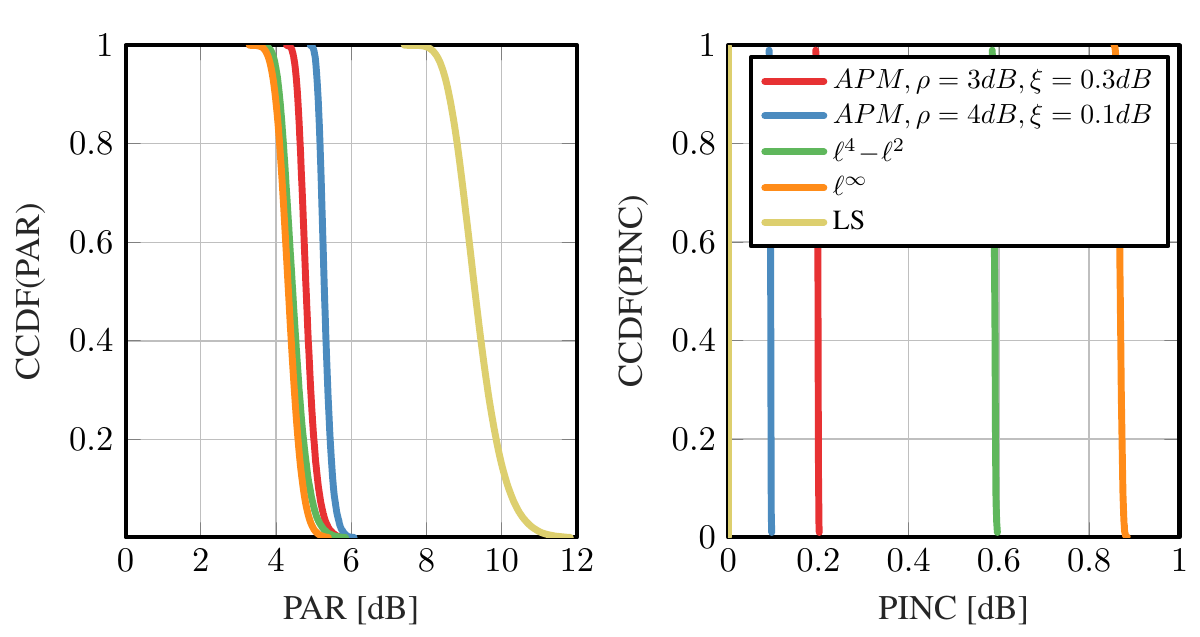}
\vspace{-0.2cm}
\caption{CCDF of PAR and PINC at the $5$th iteration of the considered JPP algorithms and of  a baseline LS precoder.}
\label{fig:ccdf_iter5}
\end{figure}


\section{Conclusions}

We have proposed a novel formulation for finding min-PAR solutions to an underdetermined linear system using an alternating projections method (APM).
Our method simultaneously bounds the PAR  of the solution vectors and their power increase (PINC), which leads to solutions that are, in general, unattainable by $\ell^\infty$-norm minimization and have lower PINC compared to the recent $\ell^p\!-\!\ell^q$-norm formulation in~\cite{taner2021ellpellqnorm}.
We have applied our approach to  a massive MU-MIMO-OFDM scenario, which has shown that the proposed APM is able to 
decrease the 99th percentile PAR by $5$\,dB in only $5$ {algorithm} iterations while keeping the PINC below 0.2\,dB. 
 

\appendices

\section{Derivation of the Projection in \fref{sec:projD}}
\label{app:KKTproof}

As a result of the PAR definition in~\fref{eq:defPAR}, we can express the PAR constraint of the set $\setD$ in \fref{eq:setD} as $|{x}_i|^2 \leq \alpha\|\vecx\|_2^2,\forall i$, where $\alpha \triangleq \rho/N$ with $1/N\leq\alpha\leq 1$.
Furthermore, since $\PINC(\vecx) \define \|\vecx\|_2^2 /  \|\vecx^\text{LS}\|_2^2 $, we can express the PINC constraint of the set $\setD$ as $\|\vecx\|_2^2 \leq P$ with  $P\triangleq \xi \|\vecx^\text{LS}\|_2^2$.
Hence, the optimization problem we wish to solve is as follows: 
\begin{align} \label{eq:proxpar_pow}
\proj{\setD}{\vecz}=\!\!\! \,\, \left\{\begin{array}{ll} 
\!\underset{\tilde\vecx\in\complexset^N}{\text{arg\,min}} &\|\vecz - \tilde\vecx\|^2_2 \\
\!\text{subject to} \!\!\!\! &  |\tilde{x}_i|^2 \!\leq \!\alpha\|\tilde\vecx\|^2_2,\forall i,  \|\tilde\vecx\|_2^2 \! \leq \!P.
\end{array}\right.
\end{align}
Here, we define the sets $\setD_1 \triangleq  \{\vecx\in\opC^N \mid \|\vecx\|^2_2\leq P\}$ and $\setD_2 \triangleq \{\vecx\in\opC^N \mid |{x}_i|^2 \leq\alpha\|\vecx\|^2_2,\forall i\}$, so that $\setD = \setD_1\cap\setD_2$.
We solve \fref{eq:proxpar_pow} by forming the following Lagrangian function with the dual variables $\vecu\in\opR^N$ and $v\in\opR$:
\begin{align} 
\mathfrak{L} (\tilde\vecx,\vecu, v) = &\textstyle \|\vecz - \tilde\vecx\|^2_2 +  \sum_{i=1}^Nu_i\big(|\tilde x_i|^2-\alpha\|\tilde\vecx\|^2_2\big) \notag \\
&+ v(\|\tilde\vecx\|_2^2 - P).  \label{eq:lagrangian_par}
\end{align}
Following the definitions of~\cite{kreutzdelgado2009complex} for complex-valued gradients, we arrive at the following partial derivatives:  
\begin{align} 
\nabla_{\tilde x_i}{\mathfrak{L} (\tilde\vecx,\vecu, v)} 
& \textstyle =\bigg(u_i + v + 1-\alpha\sum_{j=1}^Nu_j\bigg)\tilde x_i - z_i.
\end{align}
Let $t\triangleq 1-\alpha\sum_{j=1}^Nu_j$ and $\vecx=\proj{\setD}{\vecz}$ be a minimizer of~\fref{eq:proxpar_pow}. Then, the vector $\vecx$ must satisfy the KKT conditions:
\begin{enumerate}[(C\arabic*),leftmargin=*]
\item 
$\nabla_{x_i}{\mathfrak{L} (\vecx,\vecu, v)} = 0$ $\implies (u_i+v + t)x_i = z_i,\forall i.$
\item 
$u_i(|x_i|^2-\alpha\|\vecx\|^2_2)=0,\, \forall i$, and $v(\|\vecx\|_2^2 - P) = 0$.
\item 
$|x_i|^2 \leq \alpha\|\vecx\|^2_2,\,\forall i$, and $\|\vecx\|_2^2 \leq P $.
\item 
$u_i\geq 0,\,\forall i$, and $v \geq 0$.
\end{enumerate}
We note that neither the PAR nor the power constraint depends on the phases of the entries in $\vecx$. 
Hence, the entries of a minimizer $\vecx$ must match the phases of the entries of the input~$\vecz$, which implies that $u_i + v + t\geq 0$ must hold.

Let $P'\triangleq \|\vecx\|_2^2$. We define the set $\setI \triangleq \{ i: u_i>0\}$ with $|\setI|=L$ and, thus, by definition, $\setIc \triangleq \{ i: u_i=0\}$ with $|\setIc|=N\!-\!L$; we will first derive the solution assuming that~$\setI$ is known, and then show how to determine $\setI$.
In order to satisfy~(C2), we need that $|x_i|^2=\alpha P',\,\forall i \in\setI$. Then, $\setI$ denotes the indices of the largest $L$ entries of $\vecx$ (and, hence, of $\vecz$) in magnitude. 

If $\setI$ is empty, i.e., $L=0$,  then $u_i=0,\forall i$ and $ t=1$, hence, we have $v x_i=z_i,\,\forall i$, by (C1). In order to satisfy (C3), this means that $|z_i|^2\leq \alpha\|\vecz\|_2^2$, $\forall i$. In other words, $u_i=0$, $\forall i$ if and only if the vector~$\vecz$ already satisfies $\PAR(\vecz)\leq \rho$.
Then,  $\vecx = \proj{\setD_1}{\vecz} = \min\big\{1,\sqrt{P}/ \|\vecz\|_2\big\} \, \vecz$ as it is well known.

In what follows, we assume $\|\vecz\|_\infty^2 > \alpha\|\vecz\|_2^2$ so that $L\geq 1$.
We consider the following two cases separately: Case 1, where $\vecz_\setIc = \bZero_{(N-L)\times 1}$, and Case 2, where $\vecz_\setIc\neq\bZero_{(N-L)\times 1}$.

 \subsubsection*{Case 1} 
If $\vecz_\setIc=\bZero_{N-L}$, then, 
we cannot satisfy the PAR constraint by simply scaling the entries of $\vecz$, but we have to assign nonzero values to replace the zeros of $\vecz$. 
If $z_i=0$, then any value of $x_i$ could satisfy (C1) as long as $u_i+ v + t=0$. Therefore, if $z_i=0,\forall i\in\setIc$, then $\{x_i\}_{i\in\setIc}$ is not uniquely determined by the optimization problem. 
Here, we assume that the minimizer $\vecx$ will be in the following form:  
\begin{align} \label{eq:xepsilon}
x_i = \begin{cases} 
\epsilon, & i \in \setIc \\
\frac{z_i}{u_i+ v + t},& i \in \setI,
\end{cases}
\end{align}
where $0<\epsilon \leq \sqrt{\alpha P'}  =  \frac{|z_i|}{u_i+ v + t} ,\, i\in\setI$. 
From \fref{eq:xepsilon}, we have 
\begin{align}
P' = (N-L)\epsilon^2  + L(\alpha P') 
\implies \epsilon = \textstyle\frac{\sqrt{(1-\alpha L)P'}}{\sqrt{N-L}}. \label{eq:epsilon}
\end{align}
From (C1), we have that $(u_i + v + t)\epsilon =0,\,  i\in\setIc$. As $u_i=0,  \,i\in\setIc$ by definition, this implies that  $v + t = v + 1-\alpha\sum_{i\in\setI} u_i =0$, hence $\sum_{i\in\setI} u_i =(v+1)/\alpha$.
Then, since $|x_i|=|z_i|/u_i = |z_j|/u_j,\,\forall i,j\in\setI$, we have that 
\begin{align}
\textstyle \sum_{i\in\setI} u_i &=\textstyle \sum_{i\in\setI}\frac{|z_i|}{|z_j|}u_j =  \frac{\|\vecz_\setI\|_1}{|z_j|}u_j = (v+1)/\alpha \\
\implies u_j &= \textstyle \frac{(v+1) |z_j|}{\alpha\|\vecz_\setI\|_1},\, j\in\setI. \label{eq:ui_0}
\end{align}
Therefore, for $j\in\setI$, ${P'}= |z_j|^2/({\alpha} u_j^2) = \alpha \|\vecz_\setI\|_1^2 / (v+1)^2$. 
Recall that $P' \leq P$ from (C3); hence, (C2) implies that if $P' < P$, then $v=0$.
That is, if $\alpha \|\vecz_\setI\|_1^2 < P $, then $P' = \alpha \|\vecz_\setI\|_1^2$ (and $v=0$); otherwise, $P'=P$ (and $v =  \sqrt{ \frac{\alpha} {P} }  \|\vecz_\setI\|_1 - 1 $).
Finally, $P' =  \min\big\{\alpha \|\vecz_\setI\|_1^2, {P} \big\}$, and the result is given by
\begin{align} 
x_i = \begin{cases} 
\frac{\sqrt{(1-\alpha L)P'}}{\sqrt{N-L}}& i \in \setIc\\
\frac{\sqrt{\alpha P'}}{|z_i|}z_i ,& i \in \setI.
\end{cases} \label{eq:case1resultP}
\end{align}

\subsubsection*{Case~2}

If $\vecz_\setIc\!\neq\!\bZero_{N-L}$, then ${u_i + v+ t}>0$, $\forall i$ by (C1), which allows us to write $x_i$ by scaling $z_i$ in the form below:
\begin{align} \label{eq:v_xi}
x_i = \begin{cases} \frac{z_i}{v+t}, & i \in \setIc\\
\frac{z_i}{u_i+ v + t},& i \in \setI.
\end{cases}
\end{align}
Intuitively, we divide large entries of $\vecz$ by a larger constant (compared to small entries) which reduces the dynamic range.
From~\fref{eq:v_xi}, it follows that
\begin{align}
|x_i| = \sqrt{\alpha P'} & \textstyle = \frac{|z_i|}{u_i+ v+ t} ,\,i \in \setI \label{eq:zi_normx_equiv}.
\end{align}
From \fref{eq:v_xi} and \fref{eq:zi_normx_equiv}, we have  
\begin{align} \label{eq:P_prime}
\textstyle P' = \sum_{i=1}^N |x_i|^2 &= \textstyle  \frac{1}{(v+t)^2}{\sum_{i\in\setIc}|z_i|^2} + L(\alpha P'). 
\end{align} 
Note that we must have $L<1/\alpha$, since $P'>0$ and $\vecz_\setIc\neq\bZero_{N-L}$; we will revisit this fact later. 
From~\fref{eq:zi_normx_equiv} and \fref{eq:P_prime}, we have the following:
\begin{align} 
&\textstyle  P'=\frac{|z_i|^2}{\alpha(u_i + v + t)^2} = \frac{\|\vecz_\setIc\|_2^2}{(1-\alpha L)(v+t)^2} , i \in\setI \label{eq:Pprime_zIc} \\
&\implies u_i = (\beta_i-1) (v + t) , i\in\setI, \label{eq:tuj}
\end{align}
where $\beta_i \triangleq \frac{\sqrt{1-\alpha L}|z_i|}{\sqrt{\alpha}\|\vecz_\setIc\|_2} $ for $i\in\setI$.
From \fref{eq:Pprime_zIc}, we obtain 
\begin{align}
v + t &= \textstyle  \frac{\|\vecz_\setIc\|_2 }{\sqrt{(1-\alpha L)P'}}. \label{eq:t_Pprime}
\end{align}
Using \fref{eq:zi_normx_equiv} and \fref{eq:t_Pprime}, we can rewrite \fref{eq:v_xi} as 
\begin{align}
x_i = \begin{cases} 
\frac{\sqrt{(1-\alpha L)P'}}{\|\vecz_\setIc\|_2}z_i, \, &i \in \setIc\\
\frac{\sqrt{\alpha P'} }{|z_i|} z_i, \, &i \in \setI.
\end{cases} \label{eq:xi_Pprime}
\end{align}
Now, from~\fref{eq:zi_normx_equiv}, we express $u_i$ in terms of $u_j$ as follows:
\begin{align} 
& \frac{|z_i|}{u_i + v + t} =  \frac{|z_j|}{u_j + v + t}, i,j\in\setI \\
& \implies u_j = \textstyle \frac{|z_j|}{|z_i|}u_i + (v+t)\bigg(\frac{|z_j|}{|z_i|} - 1\bigg), i,j\in\setI. \label{eq:ui_uj}
\end{align}
Using \fref{eq:ui_uj}, we write $t$ in terms of $u_i$, $i\in\setI,$ as follows:
\begin{align}\label{eq:t_first}
 t & = 1-\alpha\sum_{j\in\setI} \textstyle \bigg( \frac{|z_j|}{|z_i|}u_i + (v+ t)\bigg(\frac{|z_j|}{|z_i|} - 1\bigg) \bigg)\\
&= 1-\alpha\gamma_i u_i -\alpha(\gamma_i - L)(v+t), \label{eq:t_gamma}
\end{align}
where we defined $\gamma_i \triangleq \frac{\|\vecz_{\setI}\|_1}{|z_i|}$ for $i\in\setI$. 
Inserting \fref{eq:tuj} in \fref{eq:t_gamma} followed by simplifications, we obtain
\begin{align}
& v + t = \frac{(1+v) \|\vecz_\setIc\|_2 }{\sqrt{1 - \alpha L} (\sqrt{1 - \alpha L}  \|\vecz_\setIc\|_2 + \sqrt\alpha \|\vecz_{\setI}\|_1 )} . \label{eq:t_fin}
\end{align} 
Inserting \fref{eq:t_fin} in \fref{eq:Pprime_zIc}, we have
\begin{align}
P' 
&= \big( \sqrt{1-\alpha L}\|\vecz_\setIc\|_2 + \sqrt{\alpha}\|\vecz_\setI\|_1\big)^2 / (1+v)^2. \label{eq:Pprime_fin}
\end{align}
Recall that $P' \leq P$ from (C3); hence, (C2) implies that if $P' < P$, then $v=0$.
That is, if $\big( \sqrt{1-\alpha L}\|\vecz_\setIc\|_2 + \sqrt{\alpha}\|\vecz_\setI\|_1\big)^2 < P $, then $P' = \big( \sqrt{1-\alpha L}\|\vecz_\setIc\|_2 + \sqrt{\alpha}\|\vecz_\setI\|_1\big)^2$ (and $v=0$); otherwise, $P'=P$ (and $v =  \big( \sqrt{1-\alpha L}\|\vecz_\setIc\|_2 + \sqrt{\alpha}\|\vecz_\setI\|_1\big)^2/P - 1$).
Therefore, $P' =  \min\big\{\big( \sqrt{1-\alpha L}\|\vecz_\setIc\|_2 + \sqrt{\alpha}\|\vecz_\setI\|_1\big)^2, {P} \big\}$.
Inserting \fref{eq:Pprime_fin} in \fref{eq:xi_Pprime} yields the desired final result.

\subsubsection*{Determining the Index Set $\setI$}
So far, we have assumed that we knew $\setI$ (and thus $L$); we will now explain how to determine $\setI$.
We will first consider the constraints that the set~$\setI$ should satisfy in Case~2. (i) For the condition (C3) to hold, the following must be satisfied:
\begin{align}
\textstyle  \max_{i\in\setIc}|x_i| = \max_{i\in\setIc}|z_i|/(v+t) \leq \textstyle  \sqrt{\alpha P'}. \label{eq:maxxi_ineq}
\end{align}
Inserting \fref{eq:t_fin} and \fref{eq:Pprime_fin} in \fref{eq:maxxi_ineq} yields
\begin{align}
\textstyle  \max_{i\in\setIc}|z_i| \leq   \textstyle  \sqrt{\frac{\alpha}{1-\alpha L}}\|\vecz_\setIc\|_2.   \label{eq:zi_par_cond} 
\end{align}
(ii) From \fref{eq:zi_normx_equiv}, we have that  $u_i = \frac{|z_i|}{\sqrt{\alpha P'}} - (v+t),i\in\setI$. Then, for $u_i>0, \,\forall i \in\setI$ to hold, the following  must be satisfied:
 \begin{align}
\textstyle {\min_{i\in\setI}|z_i|} &>  (v+t) \sqrt{\alpha P'} =  \textstyle \sqrt{\frac{\alpha}{1-\alpha L}} \|\vecz_\setIc\|_2. \label{eq:uj_ineq}
\end{align}
Combining \fref{eq:zi_par_cond} and \fref{eq:uj_ineq} gives us the following:
\begin{align}
\textstyle  \max_{i\in\setIc}|z_i| \leq \sqrt{\frac{\alpha}{1-\alpha L}}\|\vecz_\setIc\|_2 < \min_{i\in\setI}|z_i|. \label{eq:setIcond}
\end{align}
Note that \fref{eq:setIcond} is also trivially satisfied if $\vecz_{\setI_c}= \bZero_{N-L}$, which corresponds to Case~1.

Now, recall that $\setI$ is empty when $\vecz$ already satisfies the PAR constraint; and note that $\setI$ indexes the entries of $\vecz$ whose magnitude is too large to satisfy the PAR constraint and will be clipped to a fixed upper bound value in the corresponding entries of $\vecx$. 
Since $\vecx$ has the minimum distance to $\vecz$ while satisfying the PAR constraint, the set $\setI$ (thus $L$) should be as small as possible while the KKT conditions are satisfied, and also defined uniquely in order to avoid an ambiguity about which entries to clip.
Therefore, we begin our search for $L$ by initializing $|\setI|\!=\!L\!=\!1$ in our solution procedure and increment $L$ by 1 until $\setI$ is determined uniquely and \fref{eq:setIcond} is satisfied.
Recall that $L<1/\alpha$ from \fref{eq:P_prime}, which implies that we will test~\fref{eq:setIcond} for at most $\floor{1/\alpha}$ values of $L$ before finding the correct choice, where $\floor{\cdot}$ denotes rounding towards $-\infty$. 
Testing for the condition in \fref{eq:setIcond} requires calculating the magnitude of the entries of $\vecz$ and sorting them once regardless of how many $L$'s  we try; hence, the complexity remains $O(N\log N)$.

Finally, we make an important observation:
Computing $\proj{\setD}{\vecz}$ is equivalent to performing $\proj{\setD_1}{\proj{\setD_2}{\vecz}}$. 
As a proof sketch, we note that (i) $\proj{\setD_2}{\vecz}$ is a special case of our derivation since $\proj{\setD}{\vecz} = \proj{\setD_2}{\vecz}$ for $P=\infty$, and (ii) it is well-known that $\proj{\setD_1}{\vecx'} = \min\big\{1,\sqrt{P}/ \|\vecx'\|_2\big\} \, \vecx'$.
In \fref{sec:projD}, we computed $\vecx' = \proj{\setD_2}{\vecz}$ in Steps 2-to-4 and $\proj{\setD_1}{\vecx'}$ in Step 5 to emphasize the separability of the PAR and PINC constraints.

\bibliographystyle{IEEEtran}

\balance

\bibliography{bib/IEEEabrv,bib/confs-jrnls,bib/publishers,bib/studer,bib/vipbib,23WCNC_bib} 

\begin{thebibliography}{10}
\providecommand{\url}[1]{#1}
\csname url@samestyle\endcsname
\providecommand{\newblock}{\relax}
\providecommand{\bibinfo}[2]{#2}
\providecommand{\BIBentrySTDinterwordspacing}{\spaceskip=0pt\relax}
\providecommand{\BIBentryALTinterwordstretchfactor}{4}
\providecommand{\BIBentryALTinterwordspacing}{\spaceskip=\fontdimen2\font plus
\BIBentryALTinterwordstretchfactor\fontdimen3\font minus
  \fontdimen4\font\relax}
\providecommand{\BIBforeignlanguage}[2]{{%
\expandafter\ifx\csname l@#1\endcsname\relax
\typeout{** WARNING: IEEEtran.bst: No hyphenation pattern has been}%
\typeout{** loaded for the language `#1'. Using the pattern for}%
\typeout{** the default language instead.}%
\else
\language=\csname l@#1\endcsname
\fi
#2}}
\providecommand{\BIBdecl}{\relax}
\BIBdecl

\bibitem{larsson14a}
E.~G. Larsson, F.~Tufvesson, O.~Edfors, and T.~L. Marzetta, ``Massive {MIMO}
  for next generation wireless systems,'' \emph{{IEEE} Commun. Mag.}, vol.~52,
  no.~2, pp. 186--195, Feb. 2014.

\bibitem{NP00}
R.~van Nee and R.~Prasad, \emph{{OFDM} for wireless multimedia
  communications}.\hskip 1em plus 0.5em minus 0.4em\relax Artech House Publ.,
  2000.

\bibitem{HL05}
S.~H. Han and J.~H. Lee, ``An overview of peak-to-average power ratio reduction
  techniques for multicarrier transmission,'' \emph{IEEE Wireless Comm.},
  vol.~12, no.~2, pp. 1536--1284, Apr. 2005.

\bibitem{rahmatallah_survey}
Y.~Rahmatallah and S.~Mohan, ``Peak-to-average power ratio reduction in {OFDM}
  systems: A survey and taxonomy,'' \emph{{IEEE} Commun. Surveys Tuts.},
  vol.~15, no.~4, pp. 1567--1592, Mar. 2013.

\bibitem{bao18}
H.~Bao, J.~Fang, Q.~Wan, Z.~Chen, and T.~Jiang, ``An {ADMM} approach for {PAPR}
  reduction for large-scale {MIMO}-{OFDM} systems,'' \emph{{IEEE} Trans. Veh.
  Technol.}, vol.~67, no.~8, pp. 7407--7418, Aug. 2018.

\bibitem{kant2021asilomar}
S.~Kant, M.~Bengtsson, B.~G{\"o}ransson, G.~Fodor, and C.~Fischione, ``Robust
  {PAPR} reduction in large-scale {MIMO-OFDM} using three-operator {ADMM}-type
  techniques,'' in \emph{Proc. Asilomar Conf. Signals, Syst., Comput.}, Oct.
  2021, pp. 616--622.

\bibitem{kant2022twc}
S.~Kant, M.~Bengtsson, G.~Fodor, B.~Göransson, and C.~Fischione, ``{EVM}
  mitigation with {PAPR} and {ACLR} constraints in large-scale {MIMO}-{OFDM}
  using {TOP-ADMM},'' \emph{{IEEE} Trans. Wireless Commun.}, vol.~21, no.~11,
  pp. 9460--9481, May 2022.

\bibitem{gokceli2022}
S.~Gökceli, T.~Riihonen, T.~Levanen, and M.~Valkama, ``Machine learning based
  tuner for frequency-selective {PAPR} reduction,'' \emph{{IEEE} Trans. Veh.
  Technol.}, pp. 1--6, 2022, early access.

\bibitem{kalinov21}
A.~Kalinov, R.~Bychkov, A.~Ivanov, A.~Osinsky, and D.~Yarotsky, ``Machine
  learning-assisted {PAPR} reduction in massive {MIMO},'' \emph{{IEEE} Wireless
  Commun. Lett.}, vol.~10, no.~3, pp. 537--541, Mar. 2021.

\bibitem{jang2021globecom}
H.~Jang, S.~Jang, Y.~Park, J.~Jung, J.~Lee, and S.~Choi, ``{SeqNet}:
  Data-driven {PAPR} reduction via sequence classification,'' in \emph{Proc.
  IEEE Global Telecommun. Conf. (GLOBECOM) Workshop}, Dec. 2021, pp. 1--6.

\bibitem{studer13a}
C.~Studer and E.~G. Larsson, ``{PAR}-aware large-scale multi-user {MIMO}-{OFDM}
  downlink,'' \emph{{IEEE} J. Sel. Areas Commun.}, vol.~31, no.~2, pp.
  303--313, Feb. 2013.

\bibitem{cha14}
H.-S. Cha, H.~Chae, K.~Kim, J.~Jang, J.~Yang, and D.~K. Kim, ``Generalized
  inverse aided {PAPR}-aware linear precoder design for {MIMO}-{OFDM} system,''
  \emph{{IEEE} Commun. Lett.}, vol.~18, no.~8, pp. 1363--1366, Aug. 2014.

\bibitem{wang14}
S.~Wang, Y.~Li, and J.~Wang, ``Convex optimization based downlink precoding for
  large-scale {M}{I}{M}{O},'' in \emph{Proc. IEEE Wireless Commun. Netw. Conf.
  (WCNC)}, Apr. 2014, pp. 218--223.

\bibitem{studer15b}
\BIBentryALTinterwordspacing
C.~Studer, T.~Goldstein, W.~Yin, and R.~G. Baraniuk, ``Democratic
  representations,'' Apr. 2015. [Online]. Available:
  \url{http://arxiv.org/abs/1401.3420}
\BIBentrySTDinterwordspacing

\bibitem{guo16}
Z.~Guo, Y.~Y{\i}lmaz, and X.~Wang, ``Transmitter-centric channel estimation and
  low-{PAPR} precoding for millimeter-wave {MIMO} systems,'' \emph{{IEEE}
  Trans. Wireless Commun.}, vol.~64, no.~7, pp. 2925--2938, Jul. 2016.

\bibitem{zayani19}
R.~Zayani, H.~Shaiek, and D.~Roviras, ``{PAPR}-aware massive {MIMO}-{OFDM}
  downlink,'' \emph{{IEEE} Access}, vol.~7, pp. 25\,474--25\,484, Feb. 2019.

\bibitem{bao16}
H.~Bao, J.~Fang, Z.~Chen, H.~Li, and S.~Li, ``An efficient {B}ayesian {PAPR}
  reduction method for {OFDM}-based massive {MIMO} systems,'' \emph{{IEEE}
  Trans. Wireless Commun.}, vol.~15, no.~6, pp. 4183--4195, Jun. 2016.

\bibitem{yao19}
M.~Yao, M.~Carrick, M.~M. Sohul, V.~Marojevic, C.~D. Patterson, and J.~H. Reed,
  ``Semidefinite relaxation-based {PAPR}-aware precoding for massive
  {MIMO}-{OFDM} systems,'' \emph{{IEEE} Trans. Veh. Technol.}, vol.~68, no.~3,
  pp. 2229--2243, Mar. 2019.

\bibitem{liu19}
T.~Liu, M.~T. Hoang, Y.~Yang, and M.~Pesavento, ``A parallel optimization
  approach on the infinity norm minimization problem,'' in \emph{Proc. Eur.
  Signal Process. Conf. (EUSIPCO)}, Sep. 2019, pp. 1--5.

\bibitem{hua2022ladmm}
L.~Hua, Y.~Wang, Z.~Lian, Y.~Su, and Z.~Xie, ``{LADMM}-based {PAPR}-aware
  precoding for massive {MIMO-OFDM} downlink systems,'' \emph{{IEEE} Trans.
  Veh. Technol.}, pp. 1--11, Sep. 2022.

\bibitem{taner2021ellpellqnorm}
S.~Taner and C.~Studer, ``$\ell^p\!-\!\ell^q$-norm minimization for joint
  precoding and peak-to-average-power ratio reduction,'' in \emph{Proc.
  Asilomar Conf. Signals, Syst., Comput.}, Oct. 2021, pp. 437--442.

\bibitem{ginat2018method}
\BIBentryALTinterwordspacing
O.~Ginat, ``The method of alternating projections,'' Sep. 2018. [Online].
  Available: \url{http://arxiv.org/abs/1809.05858}
\BIBentrySTDinterwordspacing

\bibitem{GSB14}
\BIBentryALTinterwordspacing
T.~Goldstein, C.~Studer, and R.~G. Baraniuk, ``A field guide to
  forward-backward splitting with a {FASTA} implementation,'' Nov. 2014.
  [Online]. Available: \url{http://arxiv.org/abs/1411.3406}
\BIBentrySTDinterwordspacing

\bibitem{tropp2005}
J.~A. Tropp, I.~S. Dhillon, R.~W. {Heath Jr.}, and T.~Strohmer, ``Designing
  structured tight frames via an alternating projection method,'' \emph{IEEE
  Trans. Inf. Theory}, vol.~51, no.~1, pp. 188--209, Jan. 2005.

\bibitem{3gpp19a}
3GPP, ``{5G}; {NR}; base station ({BS}) radio transmission and reception,'' May
  2019, {TS} 38.104 version 15.5.0 Rel.~15.

\bibitem{studer16a}
C.~Studer and G.~Durisi, ``Quantized massive {MU-MIMO-OFDM} uplink,''
  \emph{{IEEE} Trans. Commun.}, vol.~64, no.~6, pp. 2387--2399, Jun. 2016.

\bibitem{kreutzdelgado2009complex}
\BIBentryALTinterwordspacing
K.~Kreutz-Delgado, ``The complex gradient operator and the {CR}-calculus,''
  Jun. 2009. [Online]. Available: \url{https://arxiv.org/abs/0906.4835v1}
\BIBentrySTDinterwordspacing

\end{thebibliography}
\balance
	
\end{document}